\documentclass[12pt]{article}
\usepackage{amsmath,amsthm,amssymb,euscript,epsf,epsfig}
\usepackage{array}

\newcommand{\be}{\begin{equation}}
\newcommand{\ee}{\end{equation}}
\newcommand{\bea}{\begin{eqnarray}\displaystyle}
\newcommand{\eea}{\end{eqnarray}}
\newcommand{\nn}{\nonumber}

\newcommand{\pd}{\partial}

\makeatletter
\@addtoreset{equation}{section}
\makeatother

\def\one{{\hbox{ 1\kern-.8mm l}}}
\def\zero{{\hbox{ 0\kern-1.5mm 0}}}
\def\hr{\hat{r}}

\def\hs{\hat{s}}
\def\tro{{\tilde{r}_1}}

\def\a{\alpha}      
\def\b{\beta}       
\def\g{\gamma}  \def\G{\Gamma}  
\def\d{\delta}    
\def\e{\epsilon}

\def\k{\kappa}
 
\def\m{\mu} \def\n{\nu}

\def\p{\pi} 

\def\s{\sigma}  \def\S{\Sigma}
\def\t{\tau}
  \def\vth{\vartheta}

\def\z{\zeta}

\def\nnm{ \nonumber }
\def\rp{ {r^{\prime}} }
\def\tr{ { \tilde {r } } }
\def\ts{ { \tilde {s } } }

\begin{document}
{}~
{}~
\hfill\vbox{\hbox{QMUL-PH-05-06}
}\break

\vskip .6cm

\centerline{{\Large \bf  Large-small dualities  between periodic }}
\centerline{{ \bf \Large  collapsing/expanding  branes and brane funnels }}

\medskip

\vspace*{4.0ex}

\centerline{\large \rm
C. Papageorgakis and S. Ramgoolam ${}^{\dagger}$}

\vspace*{4.0ex}
\begin{center}
{\large Department of Physics\\
Queen Mary, University of London\\
Mile End Road\\
London E1 4NS UK\\
}
\end{center}

\vspace*{5.0ex}

\centerline{\bf Abstract} \bigskip
 We consider space and time dependent fuzzy spheres
 $S^{2p}$  arising in $D1-D(2p+1)$ intersections in IIB string theory
 and  collapsing D(2p)-branes in IIA string theory.
 In the case of $S^2$, where the periodic space and time-dependent
 solutions can be described by Jacobi elliptic functions, there
 is a duality of the form   $r$ to ${1 \over r}$
 which relates the space and time dependent solutions.
 This duality is related to complex multiplication properties of the
Jacobi elliptic functions. For $S^4$ funnels, the description
of the periodic  space and time dependent solutions involves the Jacobi
Inversion problem on a  hyper-elliptic Riemann surface of genus $3$.
Special symmetries of the Riemann surface allow the reduction of
the problem to one involving a product of genus one surfaces.
The symmetries also allow a generalisation of the $r$ to ${ 1 \over r } $
duality. Some of these considerations extend to the case of the fuzzy $S^6$.

\vfill
\begin{flushright}
{\it ${}^{\dagger}${\{c.papageorgakis , s.ramgoolam \}@qmul.ac.uk}\\
}
\end{flushright}
\eject

\section{Introduction}

 Fuzzy spheres of two, four, six dimensions
 arise in a variety of related contexts. On the one
 hand they  describe the cross-sections of
 fuzzy funnels appearing at the intersection
 of D1-branes with D3, D5 or D7-branes of Type IIB
 string theory \cite{myers,nonabelian,cook}.
 In this context it is of interest
 to follow the spatial evolution of the size $r$   of the fuzzy sphere
 as a function of the co-ordinate $\sigma$  along the D-string. At the location
 of the higher dimensional brane, the cross-section of
 the funnel blows up. These equations for the funnel which arise
 either from the D-string or the $D(2p+1)$-brane worldvolume,
 can be generalised to allow for time dependence as well as spatial dependence.
 The purely time-dependent solutions are also relevant to
 the case of  spherical bound states of $D0$ and
 $D2p$-branes of Type IIA string theory.

 In the case of the fuzzy 2-sphere, there are  purely
 spatial  and  purely time-dependent solutions
 described in terms of Jacobi elliptic functions.
 The spatial and time-profiles are closely related and
 the relation follows from an $ r \rightarrow { 1 \over r } $
 duality. It is natural to introduce a complex variable $u_1 = \sigma - i t $.
 For solutions described in terms of elliptic functions,
 the inversion symmetry
 is related to the property of complex multiplication
$ u_1 \rightarrow i u_1$.
 The periodic spatial solutions describe a configuration of alternating branes and anti-branes.
 At the location of the brane or antibrane, the radius
$r$ of the funnel blows up. This is a well-understood blow-up,
expected from the geometry of a 1-brane forming a 3-brane.
The periodic solutions in time describe collapse
 followed by expansion of 2-branes. The collapse point is
 a priori a much more mysterious point, where the size of the
 fuzzy sphere is sub-stringy. Nevertheless the $r\rightarrow{1 \over r }$
 duality following from the equations of the Born-Infeld action
 imply that the zeroes of the time evolution are directly related to
 the blow-up in the spatial profile.

 In the case of the fuzzy 4-sphere, the functions
 defining the dynamics are naturally related to
 a genus $3$  hyper-elliptic curve. Using the
 conservation laws of the spatial or time evolution,
 the time elapsed or distance along the $1$-brane
 can be expressed in terms of an integral of a holomorphic
 differential on the genus $3$ hyper-elliptic curve.
 The upper limit  of the integral is the radius $r$. Inverting the integral
to express $r$ in terms of $u_1 = \sigma - i t $ is a problem
 which can be related to the Jacobi Inversion problem, with
 a constraint. Because of the symmetries of the genus $3$ curve, it
 can be mapped holomorphically to a genus $1$ and a genus $2$ curve.
 The genus $2$ curve can be further mapped to a pair of genus $1$ curves.
 The Jacobi inversion problem expressed in terms of the genus
 $2$ variables requires the introduction of a second complex variable
 $u_2$ and we find that there is a constraint which relates $u_2$ to $u_1$.
 As a result, an implicit solution to the constrained Jacobi inversion
problem can be given in terms of ordinary (genus $1$) Jacobi
 elliptic functions.  The solution is implicit in the sense that
the constraint involved is transcendental and is given in terms of elliptic functions. We give several checks
  of this solution, including a series expansion
 and  calculations of the time of collapse or distance to
 blow-ups. The symmetries which allow the reduction of the problem
 to one involving lower genus Riemann surfaces also provide
 dualities of the type $ r\rightarrow{ 1 \over r }$ which relate
 poles to zeroes.

 In section 5  we extend some of these discussions to the fuzzy 6-sphere. The space and time dependence are related to integrals of
 a holomorphic differential on a genus $5$ Riemann surface. A simple
transformation relates the problem to genus $3$. But we have not found
 a further reduction to genus one. The solution $ r ( t , \sigma )$
 can still be related to a constrained Jacobi inversion
 problem, which can be solved in terms of genus $3$ Riemann theta functions.
 As far as large-small symmetries are concerned, the
 story is much the same as for $S^4$  in the limit of large
 `initial' radius $r_0$. In the time-dependent problem by 
`initial' radius we mean
 the point where the radial velocity is zero. In the spatial problem, it is the place where the $ { dr \over d\sigma } =  0  $. For general $r_0$
 there are still inversion symmetries of the type $ ( 1 + r^4 ) \rightarrow ( 1 + r^4 )^{-1}$, but they involve fourth roots when expressed in terms of $r$,
so are not as useful.  

Appendix A.1 gives a discussion of the BPS condition for the D1-D3 
system in a  Lorentz invariant form appropriate for
space and time dependence.
 We discuss some boosts of solutions described in section 2, 
as well as the relation between the BPS equation, the Yang-Mills 
equation and the DBI equation. Appendix A.2 uses the Chern-Simons 
terms in the D1 and D3 action, to show that the solutions 
we have considered do indeed carry both $D1$ and $D3$ charge. 
Appendix B describes some Lagrangians related to the ones that appear
in the $D1-D(2p+1)$ system, and which give rise to the 
equations of motion related to higher holomorphic differentials
for the Riemann surfaces mentioned above. Appendix C describes the 
derivation of the Jacobi-$Cn$ solution for the fuzzy $S^2$ by steps
using Weierstrass $\wp$ functions, since the discussion of
section 5 on the fuzzy $S^6$ is expressed  in terms
of higher genus generalisations of the  $\wp$ functions.

\section{ Space and Time-Dependent Fuzzy $S^2$ }

\subsection{Non-abelian DBI description of non-static $D1\perp D3$ funnels }
The \emph{static} system consisting of a set of $N$ D-strings ending
on an orthogonal D3 has been thoroughly studied \cite{myers,nonabelian}.
 There exist two dual descriptions of the intersection at
 large-$N$, one from the D1 and one from the D3 worldvolume
 point of view. In the D1-picture it is described
 as a funnel of increasing radius as we approach the D3 brane, 
where the D-strings  expand into a fuzzy-$S^2$.
 In the D3-picture the worldvolume solution
 includes a  BPS magnetic
 monopole and   the Higgs field is interpreted as a
 transverse spike. Although the
 D1 picture is valid far from the D3 and
the D3 picture close to it, there is a significant region of overlap which
validates the duality. Here we will enlarge this discussion
 by lifting the static condition.
\par
We  begin by considering the non-abelian DBI action of $N$ D-strings in a flat background and with the gauge fields set to zero
\begin{equation}\label{d1dbi}
S_{DBI}^{D1}=-T_{1}\int d^{2}\sigma\; STr\sqrt{-det\left(\eta_{ab}+\lambda^{2}\partial_{a}\Phi^{i}Q_{ij}^{-1}\partial_{b}\Phi^{j}\right)\; det(Q^{ij})}\;,
\end{equation}
where $a,b$ worldvolume indices, the $\Phi$'s are worldvolume scalars,
 $\lambda\equiv2\p \a'=2\p\ell_s^2$ and
\begin{equation}
Q^{ij}=\delta^{ij}+i\lambda [\Phi^{i},\Phi^{j}]\;.
\end{equation}
The expansion of this to leading order in $\lambda$ yields the action
\begin{equation}
S_{DBI}^{D1}\simeq -T_{1}\int d^{2}\sigma \left(
N+\frac{\lambda^{2}}{2} STr
\left(\partial^{a}\Phi^{i}\partial_{a}\Phi^{i}+\frac{1}{2}
[\Phi^{i},\Phi^{j}] [\Phi^{j},\Phi^{i}]\right)+\ldots\right)
\end{equation}
and the following equations of motion at lowest order, which are the Yang-Mills equations
\begin{equation}\label{d1eom}
\partial^{a}\partial_{a}\Phi^{i}=[\Phi^{j},[\Phi^{j},\Phi^{i}]]\;.
\end{equation}
We will consider the space-time dependent ansatz
\begin{equation}\label{ansatzd1}
\Phi^{i}=\hat{R}(\sigma,\t)\;\alpha^{i},\qquad i=1,2,3\;
\end{equation}
where the $\alpha^{i}$'s are generators of the irreducible $N\times N$ matrix
 representation of the $SU(2)$ algebra
\begin{equation}
[\alpha^{i},\alpha^{j}]=2i\epsilon_{ijk}\alpha^{k}\;,
\end{equation}
with quadratic Casimir $\sum_{i=1}^3 (\a^i)^2=c \mathbf{1}_{N\times N}=(N^2-1)\mathbf{1}_{N\times N}$.
The resulting scalar field configuration describes a non-com\-mu\-ta\-tive fuzzy $S^2$ with physical radius
\be\label{rsphys}
R^2_{ph}(\s,\t)=\frac{\lambda^2}{N}Tr[\Phi^i (\s,\t)\Phi^i (\s,\t)] =
 { \lambda^2 c \over N } { \hat{R} }^2 \;.
\ee
By replacing the ansatz (\ref{ansatzd1})
 into (\ref{d1dbi}) we get the non-linear action
\begin{equation}\label{d1}
S=-T_{1}\int d^{2}\s\; STr \sqrt{1+\lambda^2 c \hat{R}^{\prime 2}-\lambda^2
c \dot{\hat{R}}^{2}}\sqrt{1+4\lambda^{2}c\hat{R}^{4}}\;.
\end{equation}
By varying this with respect to
$\hat{R}$ we recover the full equations of motion. Ignoring
corrections that  come from the application of the
symmetrised-trace prescription,  which are subleading at large $N$ \cite{rst},
 these are given by
\begin{equation}\label{nonlin}
2\lambda^{2} c \dot{\hat{R}}\hat{R}^{\prime}\dot{\hat{R}}^{\prime}+\hat{R}^{\prime\prime}(1-\lambda^{2}c\dot{\hat{R}}^{2})-
\ddot{\hat{R}}(1+\lambda^{2}c\hat{R}^{\prime
  2})=8\hat{R}^{3}\left(\frac{1+\hat{R}^{\prime
    2}\lambda^{2}c-\dot{\hat{R}}^{2}\lambda^{2}c}
{1+4\lambda^{2}c\hat{R}^{4}}\right)\;.
\end{equation}

We can convert the above formula to dimensionless variables by considering the re-scalings
\be
r=\sqrt{2\lambda\sqrt{c}}\hat{R}\; ,\qquad \tilde{\t}=\sqrt{\frac{2}{\lambda\sqrt{c}}}\t, \qquad\tilde{\s}=\sqrt{\frac{2}{\lambda\sqrt{c}}}\s\;,
\label{defdimless}
\ee
which will then imply
\bea
&& r^4=4\lambda^2 c \hat{R}^4,\quad\left(\frac{\pd{r}}{\pd\tilde{\t}}\right)^2=\lambda^{2}c\left(\frac{\pd{{\hat{R}}}}{\pd{\t}}\right)^2\; ,\quad \left(\frac{\pd{r}}{\pd\tilde{\s}}\right)^2=\lambda^{2}c\left(\frac{\pd{{\hat{R}}}}{\pd{\s}}\right)^2 \:, \\
&&\nn \frac{\pd^2 r}{\pd\tilde{\t}^2}= \frac{1}{4}\left(\frac{\pd^2 \hat{R}}{\pd{\t}^2}\right)(4\lambda^{2}c)^{3/4}\; ,\quad \frac{\pd^2 r}{\pd\tilde{\s}^2}= \frac{1}{4}\left(\frac{\pd^2 \hat{R}}{\pd{\s}^2}\right)(4\lambda^{2}c)^{3/4}\;
,\quad \frac{\pd^2 r}{\pd\tilde{\s}\pd\tilde{\t}}= \frac{1}{4}\left(\frac{\pd^2 \hat{R}}{\pd{\s}\pd{\t}}\right)(4\lambda^{2}c)^{3/4}\;.
\eea
The simplified DBI equations of motion can then be written in a
Lorentz-invariant form
\be\label{nonlinlor}
\pd_{\m}\pd^{\m}r+(\pd_{\m}\pd^{\m}r)\;(\pd_{\n}r)\;(\pd^{\n}r)-(\pd_{\m}
\pd^{\n}r)\;(\pd_{\n}r)\;(\pd^{\m}r)=2\;r^{3}\left(\frac{1+(\pd_{\m}r)\;
(\pd^{\m}r)}
{1+r^{4}}\right)\;,
\ee
where  $\m$ and $\n$ can take the values $\tilde{\s},\tilde{\t}$. Further
aspects of Lorentz invariance and boosted solutions are discussed in 
Appendix A. 
\par
One can also write down
 the re-scaled action and  energy density of the configuration,
 by making use of the dimensionless variables with
 the dots and primes implying differentiation with respect to
the re-scaled time and space respectively
\bea\label{en}
\nn
\tilde S _3&=&-\int d^2\s\sqrt{(1+r'^2-\dot r ^2)(1+r^4)}\\
E&=& (1+r^{\prime 2})\frac{\sqrt{1+r^{4}}}{\sqrt{1+r^{\prime 2}-\dot{r}^{2}}}
\;.
\eea

\par
We will now switch to the dual picture. The abelian DBI action for a
 $D3$-brane with a general gauge field, a single transverse scalar
 in the $x^9$ direction, in a flat background is
\begin{equation}
S_{DBI}^{D3}=-T_{3}\int d^{4}\s\sqrt{-det(\eta_{ab}+\lambda^{2}\partial_{a}
\Phi\partial_{b}\Phi+\lambda F_{ab})}\;.
\end{equation}
The determinant, for a general gauge field, can be calculated and gives
\begin{eqnarray}\label{dyonextra}
\nonumber S_{DBI}^{D3} &=& -T_{3} \int d^{4}\s \sqrt{1
  +\lambda^{2}|\vec{B}|^{2}+\lambda^{2}(\vec{\nabla}\Phi)^{2}+\lambda^{4}(\vec{B}\cdot\vec{\nabla}\Phi)^{2}-\lambda^{2}|\vec{E}|^{2}-\lambda^{2}
\dot{\Phi}^{2}}\\
& &
  \overline{-\lambda^{4}|\vec{E}\times\vec{\nabla}\Phi|^{2}-\lambda^{4}\dot{\Phi}^{2}|\vec{B}|^{2}-\lambda^{4}(\vec{E}\cdot\vec{B})^{2}
+2\lambda^{4}\dot{\Phi}\;\vec{\nabla}\Phi
\cdot(\vec{B}\times\vec{E})}\;,
\end{eqnarray}
from which one can derive the spherically symmetric equations of motion,
 in the absence of electric fields and with a single, radial
component for the magnetic field on the $D3$, $\vec B = \mp \frac{N}{2 R_{D3}}\hat R_{D3}$
\begin{equation}\label{phis}
\lambda\Phi^{\prime\prime}(1-\lambda^{2}\dot{\Phi}^{2})-\lambda\ddot{\Phi}(1+\lambda^{2}
\Phi^{\prime 2})+2\lambda^{3}\Phi^{\prime}\dot{\Phi}\dot{\Phi}^{\prime}=-8R_{D3}^{3}
\lambda\Phi^{\prime}
\frac{(1+\lambda^{2}\Phi^{\prime 2}-\lambda^{2}\dot{\Phi}^{2})}{(4R_{D3}^4+\lambda^{2}N^{2})}\;.
\end{equation}
Note that for this configuration, the field $\Phi$ only depends on the D3 
radial co-ordinate $R_{D3}$ and time $t$.
Expression (\ref{phis}) then looks similar to  (\ref{nonlin}) and we can 
show that it is indeed the same, when written in
terms of $D1$ world-volume quantities. We consider total differentials of 
the fuzzy sphere physical radius $R_{ph}(\s,\t)$
\begin{equation}
dR_{ph}=\frac{\partial R_{ph}}{\partial \s}\Big{|}_{\t} d\s+\frac{\partial
R_{ph}}{\partial \t}\Big{|}_{\s}d\t
\end{equation}
and recover for constant $R_{ph}$ and $\t$ respectively
\begin{equation}\label{rphi}
\frac{\partial \s}{\partial \t}\Big{|}_{R_{ph}}=-\frac{\dot{R}_{ph}}
{R_{ph}^{\prime}}=\lambda\dot{\Phi}\;,\qquad\frac{\partial \s}{\partial R_{ph}}
\Big{|}_{\tau}=\frac{1}{R_{ph}^{\prime}}=\lambda\Phi^{\prime}\;,
\end{equation}
where we are making use of the identifications\footnote{See also Appendix A.}
 $R_{ph}=R_{D3}, \t=t$ and $\s=\lambda\Phi$.
Then the second order derivatives of $\s$ are;
\begin{equation}
\frac{\partial}{\partial R_{ph}}\Big{|}_{\t}\frac{\partial \s}{\partial \t}\Big{|}_{R_{ph}}=\lambda\dot{\Phi}^{\prime}\;,\qquad\frac{\partial}{\partial \t}\Big{|}_{R_{ph}}\frac{\partial \s}{\partial  \t}\Big{|}_{R_{ph}}=\lambda\ddot{\Phi}\qquad\textrm{and}\qquad\frac{\partial}{\partial R_{ph}}\Big{|}_{\t}\frac{\partial \s}{\partial R_{ph}}\Big{|}_{\t}=\lambda\Phi^{\prime\prime}\;.
\end{equation}
The D1-brane solution $ R_{ph} ( \sigma , t ) $ can be inverted to give
$ \sigma ( R_{ph} , t ) $.  By employing the following relations
\begin{equation}
\frac{\partial f \left( \s (R_{ph},t),t\right)}{\partial t}\Big{|}_{R_{ph}} =\frac{\partial f}{\partial t}\Big{|}_{\s}
+\frac{\partial f}{\partial \s}\Big{|}_{t}\frac{\partial\s}{\partial t}\Big{|}_{R_{ph}}\quad\textrm{and}\quad\frac{\partial f \left( \s (R_{ph},t),t\right)}{\partial R_{ph}}\Big{|}_{t}=\frac{\partial f}{\partial\s}\Big{|}_{t}\frac{\partial\s}{\partial R_{ph}}\Big{|}_{t}\;,
\end{equation}
we have
\bea
\nn\lambda\Phi^{\prime\prime}=-\frac{R_{ph}^{\prime\prime}}{R_{ph}^{\prime 3}}\;,&&\quad\lambda\ddot{\Phi}=-\frac{1}{R_{ph}^{\prime 2}}\left(R_{ph}^{\prime}\ddot{R}_{ph}-2\dot{R}_{ph}\dot{R}_{ph}^{\prime}+\dot{R}_{ph}^{2}
\frac{R_{ph}^{\prime\prime}}{R_{ph}^{\prime}}\right)\\
\textrm{and}&&\quad\lambda\dot{\Phi}^{\prime}=-\frac{1}{R_{ph}^{\prime 2}}\left(\dot{R}_{ph}^{\prime}-\dot{R}_{ph}\frac{R_{ph}^{\prime\prime}}{R_{ph}^{\prime}}\right)\;.
\eea
By replacing these into (\ref{phis}), one recovers the exact non-linear equations of motion (\ref{nonlin}) in terms of the physical radius $R_{ph}$. This guarantees that any space-time dependent
solutions of (\ref{nonlin}) will have a corresponding
 dual solution on the $D3$ side.
\subsection{ Arrays of branes in space and Collapse/Re-expansion in
time dependence}
We  now   restrict to purely time dependent solutions of
 equation (\ref{nonlinlor}). The  resulting $DBI$
equations of motion  are identical to those coming from a
 Lagrangian  which describes a set of $N$ D0's,
 expanded into a fuzzy $S^2$.
 This configuration also has an equivalent dual DBI description is in
 terms of a spherical D2-brane with $N$-units of magnetic flux \cite{rst}.
 To simplify the  notation,  the re-scaled variables
$ \tilde \tau , \tilde \sigma $ of (\ref{defdimless})  will be called
$t , \sigma $.  Then the  conserved
 energy density (\ref{en}) (or energy in the D0-D2 context) at large $N$
  is
\be
E = { \sqrt { 1 +   r^4 } \over \sqrt { 1 -  { \dot r }^2 } }\;.
\ee
 If  $r_0$ is the initial radius of the collapsing configuration where  $ \dot r = 0 $, $ E = \sqrt { 1 + r_0^4 } $
 and we get
\be
\dot{r}^2 =  { { r_0^4 - r^4 } \over { 1 + r_0^4  }}\;.
\label{seq}
\ee
 This allows us to write
\be
\int_0^t dt = \int_{r_0}^{r}  { \sqrt { 1 + r_0^4 }  \over \sqrt { r^4 - r_0^4 } }\;,
\ee
which can be inverted to give
\begin{equation}\label{collapse}
r(t)=\pm r_{0}Cn\left(\tilde t ,\frac{1}{\sqrt{2}}\right)\;,
\end{equation}
where $ \tilde t = \frac{\sqrt{2} r_{0}
t}{\sqrt{r_{0}^{4}+1}}$. Such solutions were first  described in
\cite{coltuck} and more recently in
\cite{rst,kata,CLI,CLII}\footnote{See Appendix C for
a derivation of this result using Weierstrass $\wp$-functions.}.

\par
The function $r(t)$ describes a D2-brane  of radius starting at $r=r_0$,
at $t=0$. It decreases to zero, then goes negative down to a minimum
$-r_0$ and then increases back through zero to the initial position.
The cycle is then repeated (see Figure 1). The region of negative
$r$ is somewhat mysterious, but we believe the correct interpretation
follows if, elaborating on (\ref{rsphys}), we define the physical radius as
\be
R_{ph} = + { \lambda \over \sqrt {N}  }  \sqrt { Tr ( \Phi_i^2 ) }
         = { \sqrt { \lambda } c^{1 \over 4 } r  \over N \sqrt{2}  }
\ee
in the region of positive $r$  and as
\be
 R_{ph} = -  { \lambda \over \sqrt {N}  }  \sqrt { Tr ( \Phi_i^2 ) }
        = - { \sqrt { \lambda } c^{1 \over 4 } r  \over N \sqrt{2}  }
\ee
in the region of negative $r$.
This guarantees that $R_{ph}$ remains positive. The change in sign
at $0$ should not be viewed as a discontinuity that invalidates the
use of the derivative expansion in the Dirac-Born-Infeld action,
since the quantity that appears in the action is $r$ (or  $ \hat R $)
 rather than $R_{ph}$.
Continuity of the time derivative $\partial_t r $ at $r=0$ also guarantees
that the $D3$-brane (or $D2$-brane) charge is continuous\footnote{See Appendix A for expressions for the charge.}. This interpretation is compatible
 with the one in \cite{myers}, where  different signs of the ${ \hat R }$
were interpreted as corresponding to either a brane or an
anti-brane emerging at the blow-up of the $S^2$ funnel.

Instead of dropping space dependence we can restrict ourselves to a static problem by making our ansatz time independent.
There is a conserved pressure $T^{\sigma \sigma }$
\begin{equation}
\frac{\partial T^{\s\s}}{\partial \s}=0\;.
\end{equation}
By plugging-in the correct expression we recover
\begin{equation}\label{momentum}
\frac{\partial}{\partial\s}\sqrt{\frac{1+r^{4}}{1+r^{\prime 2}}}=0\;,
\end{equation}
which can be combined with the initial condition $r^{\prime}=0$ at $r_{0}$ to give
\begin{equation}\label{almost}
r^{\prime 2}=\left(\frac{1+r^{4}}{1+r_{0}^{4}}\right)\;.
\end{equation}
\par
The purely space-dependent and the purely time-dependent equations
are related by Wick rotation $ t \rightarrow i \sigma $. To apply  this to the
solution (\ref{collapse}) we can use the identity
\be\label{cplxmult}
 Cn\left(ix,\frac{1}{\sqrt{2}}\right)= { 1 \over Cn(x,\frac{1}{\sqrt{2}}) }\;,
\ee
which is an example of a complex multiplication formula 
\cite{Chandrasekharan}.
Therefore, the first order equation for the static
 configuration has  solutions in terms of the
 Jacobi elliptic functions
\begin{equation}\label{cn}
r(\s)=\pm r_{0}\frac{1}{Cn\left(\tilde \sigma ,\frac{1}{\sqrt{2}}\right)}\;,
\end{equation}
where $ \tilde \sigma = \frac{\sqrt{2} r_{0}
\s}{\sqrt{r_{0}^{4}+1}} $ and it can be verified that these also satisfy the full DBI equations of motion.
This solution is not BPS and does not satisfy the YM. The relations 
between the DBI, YM and BPS equations are discussed in Appendix A. 
\par
The $r(\s)$ plot reveals that it represents an infinite,
 periodic, alternating brane-anti-brane array, with $D1$-funnels extending between them.
 The values of $ \sigma $  where  $ r$ blows up, i.e. the poles of the
$Cn$-function,
correspond to locations of D3-branes and anti-D3-branes.  This follows because
the derivative  $ { \partial r \over \partial \sigma }$ changes sign
between successive poles.  This derivative appears in the computation of the D3-charge
 from the Chern-Simons terms in the D1-worldvolume.
Alternatively, we can pick an oriented set of axes
 on one brane and transport it along the funnel to the neighbouring brane
to  find that the orientation has changed.
On the left and right of a blow-up point the sign of the derivative is the same which
 is consistent with the fact that the charge of a brane measured from either the left or right
 should give the same answer.
\par
This type of solution captures the known results of F and D-strings stretching
between D3 and anti-D3's \cite{callan,gibbons} by restricting to a half-period
 of the elliptic function in the space evolution. It is also possible to
 recover the BPS configurations of \cite{nonabelian} which were obtained
 by considering the minimum energy condition of the static funnel, where $\dot{r}=0$
\be\label{nahm}
\partial_\s\Phi^i=\pm\frac{i}{2}\e^{ijk}[\Phi^j,\Phi^k]\;.
\ee
This is equivalent to the Nahm equation \cite{nahm} and is also the BPS
 condition. In dimensionless variables it translates to $r'^2=r^{4}$ and
 has a solution in terms of $r=\pm 1/(\s-\s_{\infty})$, with $\s_{\infty}$
 denoting the point in space where the funnel blows-up. We will restrict
 the general solution to a quarter-period and consider the expansion
around the first blow-up which occurs at $Cn(K(\frac{1}{\sqrt{2}}),\frac{1}{\sqrt{2}})$,
 i.e. close to the D3-brane. We get
\bea
r =\frac{r_0}{Cn\left(\frac{\sqrt{2}r_0 \s}{\sqrt{1+r_0^4}}\right)} & \simeq
- \frac{r_0}{\frac{1}{\sqrt{2}}\left(\frac{\sqrt{2}r_0\s}{\sqrt{1+r_0^4}}
-K(\frac{1}{\sqrt{2}})\right)}\cr
& =  -\frac{ \sqrt{ 1 + r_0^4 }    }{  \s  -
\frac{ \sqrt{ 1 + r_0^4 } K(\frac{1}{\sqrt{2}})}{ \sqrt{2}  r_0}}\;.
\eea
This is of the form  $ r =  -{    \sqrt{ 1 + r_0^4 }\over { \sigma - \sigma_{\infty} }} $
which goes to $  r =  -{  1 \over { \sigma - \sigma_{\infty} }}$  as $ r_0 \rightarrow 0 $.

\begin{figure}[t]\label{fig:ell}
\begin{center}
\includegraphics[height=9cm,width=16cm]{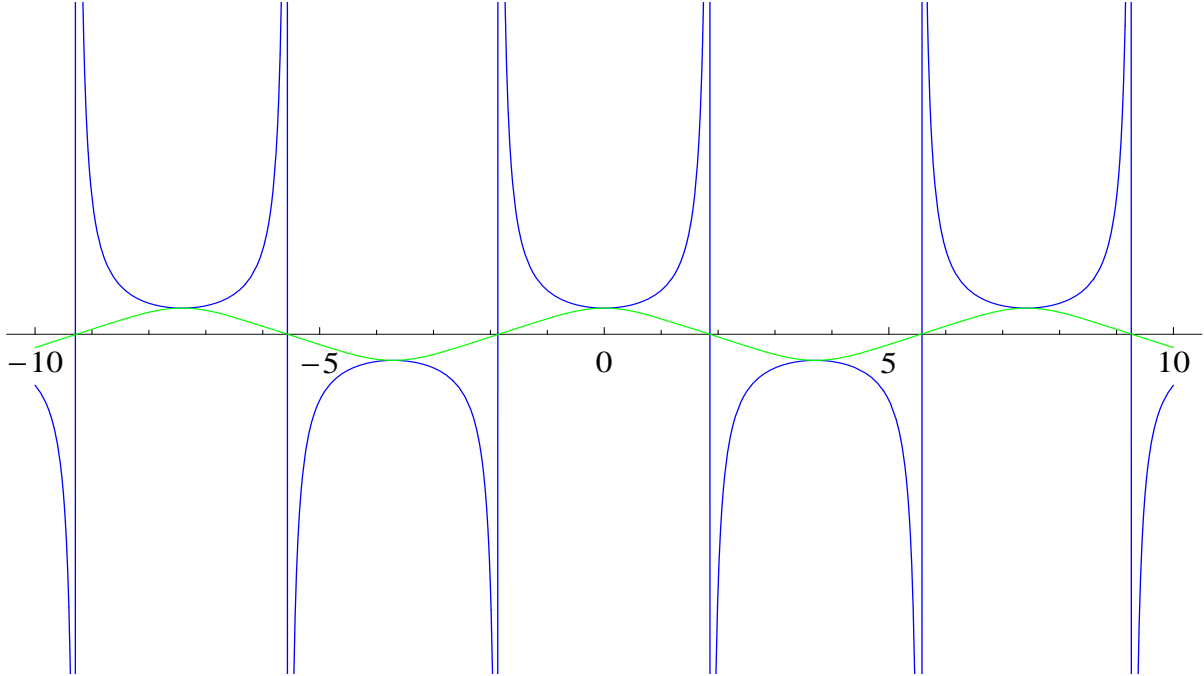}
\caption{Analytic plot of the Jacobi elliptic function solution for the static
 fuzzy-$S^2$ funnel array and the collapsing 2-sphere for $r_0=1$.}
\end{center}
\end{figure}

\subsection{ An $r \rightarrow 1/r $ duality
 between the Euclidean and Lorentzian DBI equations }

Defining $ s = { dr \over dt }$ the
 equation (\ref{seq}) can be written as
\be
s^2 = { r_0^4 - r^4  \over 1 + r_0^4 }
\label{gen1curve}
\ee
Viewing $ s$ and $r$ as two complex variables
constrained by one equation, this defines a
genus one Riemann surface.
The quantity  $ { dr \over s } $ which  gives the infinitesimal
time elapsed is an interesting geometrical quantity related to
the Riemann surface, i.e. the holomorphic differential.

 The curve (\ref{gen1curve})  has a number of automorphisms of interest.
One checks that $ R = { r_0^2 \over r } $, $ \tilde s =
{ i s r_0^2 \over r^2 } $ leaves the equation of the curve invariant.
This automorphism of the curve leads directly to the complex multiplication
identity (\ref{cplxmult}) which relates the spatial
 and time-dependent solutions.
Another, not unrelated, automorphism acts as $ R  \rightarrow
 { 1 \over r } $, $ R_0 \rightarrow { 1 \over r_0 } $,
$ \tilde s = { i s \over  r^2  }$. The relation between
$ \tilde s $ and $s$ is equivalent to a Wick rotation of the time variable. 
The transformation $ r \rightarrow { 1 \over r } $ can also be taken to
act on the second order equation since it does not involve $r_0$ (which
does not appear in the second order equation). The spatial BPS
solution to the second order equation can be acted upon by this
transformation.  The outcome is a  time-dependent
solution describing a brane collapsing at the speed of light $r=\pm(t-t_{\infty})$.
 This solution  can be derived as a $r_0 \rightarrow \infty $ limit of the general
 time-dependent elliptic solution, in much the same way as the
 BPS solution was derived as an $r_0 \rightarrow 0 $ limit of
 the spatial elliptic solutions.

 The action of the $ r \rightarrow { 1 \over r } $ transformation 
on the second order equations  can be seen explicitly.
 In the case of pure time dependence, the equation of
 motion in dimensionless variables is

\be
\ddot r = -  2r^3  \left( {   1 - \dot r^2 \over {  1 + r^4} } \right)\;,
\label{puret}
\ee

 while in the case of pure spatial dependence
\be
{r^{\prime\prime}} = 2 r^3   \left({ 1 +   \rp^2  \over 1 + r^4 } \right)\;.
\ee

 A substitution $r = 1/R $ can be used to transform (\ref{puret})
 using
\bea
\dot r &=& -{  1 \over R^2 } \dot R \nnm\\
\ddot r &=&  -{  1 \over R^2 } \ddot R + { 2 \over R^3 } {\dot R}^2 \nnm\;,
\eea

to get
\be
- { \ddot R \over R^2} + { 2 \over R^3 } ( \dot R )^2
= -{  2 \over R^3} { 1 - { { \dot R}^2\over R^4}   \over { 1 + 1/R^4} }\;,
\ee

which can be simplified to
\be
\ddot R = 2 R^3   \left({  1 + \dot R^2  \over  1  + R^4  } \right)\;.
\ee

So the effect of transforming $ r \rightarrow { 1 \over r }  $  
and renaming $ t \rightarrow \sigma $ is the same
as the substitution $ t \rightarrow i\sigma $. This explains 
the relation between (\ref{collapse}) and (\ref{cn}) which was previously 
obtained using the complex multiplication property (\ref{cplxmult}) of 
Jacobi-$Cn$ functions. 

\section{ Space and Time dependent fuzzy $S^4$ }

\subsection{Equivalence of the action for the $D1\perp D5$ intersection}
\par
We will extend the consideration of space and time
 dependent solutions to DBI for  the case of the $D1\perp D5$ intersection,
 which involves a fuzzy $S^4$, generalising the
purely spatial discussion of \cite{nonabelian}.
 The equations are also relevant to
the time-dependence of  fuzzy spherical $D0$-$D4$ systems which
 have been studied in the Yang-Mills limit in \cite{castelino}.
 On the $D1$ side, we will have five transverse scalar fields,
satisfying the ansatz
\begin{equation}\label{d5ansatz}
    \Phi^{i}(\s,\t)=\pm \hat{R}(\s,\t)G^{i}, \qquad i=1,\ldots
    5\;,
\end{equation}
where the $G^{i}$'s are given by the  action of $SO(5)$ gamma matrices
on the totally symmetric $n$-fold tensor
product of the basic spinor, the dimension of 
which is related to $n$ by
\begin{equation}\label{Nn}
    N=\frac{(n+1)(n+2)(n+3)}{6}\;.
\end{equation}
The radial profile and the fuzzy-$S^{4}$ physical radius are again
related by
\be
 R_{ph}(\s ,\t)=\sqrt{c}\lambda\hat{R}(\s,\t)\;,
 \ee
 where $c$
is the `Casimir'
$G^{i}G^{i}=c\mathbf{1}_{N}=n(n+4)\mathbf{1}_{N\times N}$.
By plugging the ansatz (\ref{d5ansatz})
 into the action and by considering the large-$N$ behaviour of the
 configuration, one gets in dimensionless variables defined just as for the $S^2$ case in (\ref{defdimless}),
\begin{equation}\label{d15}
    S_{1}=-N\;T_{1}\int d^{2}\s
    \sqrt{1+r^{\prime 2}-\dot{r}^{2}}\left(1+r^{4}\right)\;.
\end{equation}
As in the $D1\perp D3$ case, we can write the equations of motion
for this configuration in a Lorentz-invariant way. The result is
the same as before with the exception of the
pre-factor on the  right-hand-side
 \be
\pd_{\m}\pd^{\m}r+(\pd_{\m}\pd^{\m}r)\;(\pd_{\n}r)\;(\pd^{\n}r)-(\pd_{\m}
\pd^{\n}r)\;(\pd_{\n}r)\;(\pd^{\m}r)=4\;r^{3}\left(\frac{1+(\pd_{\m}r)\;
(\pd^{\m}r)} {1+r^{4}}\right)\;,
\ee
 where again $\m$ and $\n$ can
take the values $\s,\t$.

\par
Let us now look at  the $D5$ side: The world-volume action
for $n$ $D5$-branes, with one transverse scalar excited, is as
usual
\begin{equation}
    S_{5}=-T_{5}\int d^{6}\s Str\sqrt{-det(G_{ab}+\lambda^{2}\partial_{a}\Phi\partial_{b}\Phi+\lambda
    F_{ab}})\;.
\end{equation}

Introducing spherical co-ordinates with radius $R_{D5}$ and angles $\a^i$ ($i=1,\ldots,4$), we will have $ds^2=-dt^2+dR_{D5}^2+R_{D5}^2g_{ij} d\a^i d\a^j$,
where $g_{ij}$ is  the metric of a unit four-sphere, the volume of which is given by
$\int d^{4}\a\sqrt{g}=8\p^{2}/3$.
We take the construction of  \cite{nonabelian} with homogeneous instantons and, keeping the same gauge field, we generalise $ \Phi( R_{D5} ) $ to $ \Phi ( R_{D5} ,t )$. We will not review this here but just state that by using the above, we can reduce the previous expression to
\begin{equation}\label{actiond5}
    S_{5}=-T_{5}\int d^{6}\s\sqrt{g}
    \sqrt{1+\lambda^{2}\Phi^{\prime 2}-\lambda^{2}\dot{\Phi}^{2}}\left(nR_{D5}^{4}+\frac{3}{2}N\lambda^{2}\right)\;,
\end{equation}
 Then by implementing  once again the relations (\ref{rphi}) we recover
\begin{equation}
    S_{5}=-T_{5}\int d^{4}\a\wedge dt\wedge dR_{ph}\;
    \sqrt{g}\sqrt{1+\frac{1}{R_{ph}^{\prime 2}}-\frac{\dot{R}_{ph}^{2}}{R_{ph}^{\prime 2}}}\left(nR_{ph}^{4}+\frac{3}{2}N\lambda^{2}\right)\;.
\end{equation}
This can be easily manipulated to yield the following result
\begin{equation}
    S_{5}=-T_{1}N\int d\t\wedge d\s
    \sqrt{1+R_{ph}^{\prime 2}-\dot{R}_{ph}^{2}}\left(1+\frac{2n}{3N\lambda^{2}}R_{ph}^4\right)\;.
\end{equation}
Using  $\frac{2n}{3N}\sim\frac{4}{c}$, which holds
 for large-$N$ and by once again employing dimensionless variables, this becomes
\be
S_{5}=-T_{1}N\int d\t\wedge d\s
    \sqrt{1+r^{\prime 2}-\dot{r}^{2}}\left(1+r^4\right)\;,
\ee
which agrees with (\ref{d15}) and can be further simplified to the re-scaled action
\be\label{d15re}
\tilde S _5=-\int d^2 \s \sqrt{1+r'^2-\dot r ^2}(1+r^4)\;.
\ee

 Thus, every space-time dependent solution 
 described by the DBI equations of motion from the D1 worldvolume  will have an equivalent description on the D5 worldvolume.

\subsection{Solutions for the space-dependent fuzzy spheres: Funnels }

Assuming $ \dot r = 0 $, the action  (\ref{d15re}) becomes
\be
\tilde S _5 = -  \int d^2 \sigma \sqrt { 1 + {r^{\prime} }^2 } ~~( 1 + r^4 )\;.
\ee
There exists a conserved pressure $ T^{\sigma \sigma } = P $ given by
\be
P = { 1 + r^4  \over  \sqrt { 1  + \rp^2 } } \;,
\ee
which can be solved to give $ \rp   $
\bea
\left  ( { dr \over d \sigma  }\right)^2
&=& {  {1 \over P^2 } -1  } +  { ( 1  + r^4 )^2 \over P^2 }  \nnm \\
&=& {   ( 2 r^4 + r^8 ) - ( 2r_0^4 + r_0^8 ) \over  ( 1 + r_0^4 )^2 } \nnm\;.
\label{spceq}
\eea
Defining $ { dr \over d \sigma  } = s $ this can be expressed as
\be
s^2 =   { ( r^4 - r_0^4 ) ( r^4 - r_1^4 ) \over ( 1 + r_0^4 )^2 }
\label{spccurv}
\ee
The roots
in the  equation above correspond to
$ ( 1 + r^4 )= ( 1 + r_0^4) $ and $ ( 1 + r^4)  = - ( 1 + r_0^4 )$.
The second possibility gives $ r_1^4 = -2 - r_0^4  $.
By differentiating the pressure, a second order differential
equation can be derived
\bea
{ d^2 r \over d \sigma^2 } &=& { 4 r^3 ( 1 + \rp^2 ) \over ( 1 + r^4 ) } \nnm\\
    &=& { 4 r^3 ( 1 + r^4 ) \over P^2 }    \;.
\label{spcsec}
\eea
An integral formula can be written for the distance along the $D1$-brane
using (\ref{spceq})
\be
\int_0^{\sigma}  d \sigma = \int_{r_0}^{r}  dr  { ( 1 + r_0^4 )
\over \sqrt { ( 2 r^4 + r^8)-  ( 2r_0^4 + r_0^8 ) } }
\label{integsig}\;,
\ee
where we have taken the zero of $\sigma $ to be at the place $r = r_0$  where $\rp = 0 $.
We deduce from the integral that there is a finite value of $\sigma$, denoted
as $ \Sigma $, where $r$ has increased to infinity. This is similar to 
 part of the spatial solution of the $D1$-$D3$ system described by the $Cn$ function (see Figure 1).
We can show that the full periodic structure analogous to that of the spatial $D1$-$D3$
system follows in the $D1$-$D5$ system, by using symmetries of the equations
and the requirement that the derivative $ { dr \over d\sigma } $ is continuous.
Note that (\ref{spceq}) is symmetric under the operations
\bea
&  I_{\sigma} : r ( \sigma )  \rightarrow  r ( - \sigma )   \nn \\
&  I_{r} : r( \sigma)  \rightarrow - r ( \sigma )  \nn\\
&  T_{\Sigma } : r(\sigma ) \rightarrow r ( \sigma - \Sigma ) \:.
\label{symmsqual}
\eea
The branch with $r$ increasing from $r_0$ to $\infty$, as $\sigma $
changes from $0$ to $\Sigma$, can be acted on by $ T_{2 \Sigma} I_{\sigma}I_r$
to yield a branch where $r$ increases from $-\infty$ to $-r_0$ over
$ \Sigma \le \sigma \le 2 \Sigma $.
Acting with $ T_{2 \Sigma} I_r $ gives a branch where $r$ decreases
from $-r_0$ to $ - \infty$ over $  2 \Sigma  \le \sigma \le 3 \Sigma $.
Finally a transformation of the original branch by $T_{4 \Sigma}I_{\sigma} $
gives $r $ decreasing from $ \infty$ to $ r_0$  over 
$ 3 \Sigma \le \sigma \le 4 \Sigma $. These four branches
patch together without discontinuity in $r$ or ${ dr \over d \sigma}$
 and can further be translated
by integer multiples of $4\S$, to give a picture qualitatively similar
to the $D1$-$D3$ case, but now describing $D1$ funnels between
alternating $D5$ and anti-$D5$.
The profile has the properties
\bea\label{arraypol}
r ( \sigma +  2 \Sigma      ) &=& - r ( \sigma ) \nnm \\
r ( \sigma + 4 m  \Sigma   ) &=& r ( \sigma ) \nnm \\
{ 1 \over  {r ( ~  (2m+1 ) \Sigma  ~  ) } }  &=& 0    \;,
\eea
where $m$ is an arbitrary integer.
This picture, including the poles and periodicity,  will be recovered
with an improved quantitative description of $r$ as a function of
complex arguments in section 4 (see Figure 2).

\subsection{Solutions for the time-dependent fuzzy spheres: Collapsing spheres }
In the time-dependent case, there is a conserved energy
\be
E =  {  ( 1 + r^4  ) \over \sqrt { ( 1 - \dot r^2 ) }   } \;.
\ee
implying
\be
( 1 + r_0^4)^2 = { ( 1 + r^4  )^2 \over ( 1 - \dot r^2 ) }\;.
\label{eqfrts}
\ee
This can be solved for the velocity
\bea
 \left({dr \over dt}\right)^2 &=& { 1 - {1 \over E^2} } -
 { 1 \over E^2 } (  2 r^4 + r^8 )   \nnm\\
&=&   { ( 2 r_0^4 + r_0^8 ) - ( 2 r^4 + r^8 ) \over ( 1 + r_0^4 )^2 } \nnm\;.
\label{dotreq}
\eea
Writing $ s = { dr \over dt } $
\bea
s^2 &=& {(1 + r_0^4)^2 - ( 1 + r^4)^2  \over (1 + r_0^4)^2}   \\
    &=&  { - ( r^4 - r_0^4 ) ( r^4 - r_1^4 ) \over  (1 + r_0^4)^2 }
\label{eqwrts}\;.
\eea
A trivial redefinition
 $ s \rightarrow i s $ relates the equation (\ref{eqwrts})
to  (\ref{spccurv}).
The time evolved can be written in terms of the radial distance
\be
\int_0^t dt = \int_{r_0}^{r} dr { ( 1 + r_0^4 ) \over \sqrt {  ( 2 r_0^4 + r_0^8 ) -
 ( 2 r^4 + r^8 )}
 } \;.
\label{integt}
\ee
Differentiating (\ref{dotreq}) gives a second order equation
\bea
\ddot r &=& { - 4 r^3 ( 1 + r^4 ) \over E^2 } \nnm \\
        &=& { - 4 r^3 ( 1 + r^4 ) \over  ( 1 + r_0^4 )^2  }
\label{timsec} \;.
\eea
We can see from (\ref{integt}) that the time taken to start
from $r=r_0$ and reach $r= 0$ is finite. We will call this
finite time interval the time of collapse $T$.
As in the spatial problem there are symmetries of the equation
\bea
&  I_{t} : r ( t )  \rightarrow  r ( - t )   \nn \\
&  I_{r} : r( t )  \rightarrow - r ( t )  \nn\\
&  T_{T } : r(t ) \rightarrow r ( t - T ) \;.
\label{tsymmsqual}
\eea
Following the same steps as in the spatial case, we can act successively
with $ I_{2\Sigma} I_t I_r $ , $ I_{2 \Sigma }I_r $, $ I_{3 \Sigma }I_{t} $
and then with integer multiples of $ 4T$ to produce a periodic solution
defined for positive and negative time. As discussed before and in analogy to the $D1$-$D3$ scenario,
this can be interpreted in terms of collapsing-expanding $D4$-branes.
The radius as a function of $t$ has the properties
\bea\label{arrayzer}
r ( t + 4m T ) &=& r ( t ) \nn \\
r ( T ) = r (~  ( 2 m +1 ) T ~ ) &=& 0 \nn \\
r ( t + 2mT ) &=& - r ( t )  \;,
\eea
where $m$ is an arbitrary integer.
We will see in the following that it will be useful to define
a complex variable $u_1$ whose real part is related to
$\sigma $ and whose imaginary part is related to $t$
as $ u_1 = \sigma - i t $.
Unlike the case of the fuzzy two-sphere, we are not dealing
simply with a variable $u$ living on a torus. Rather it will
become necessary to introduce a second complex variable
$u_2$ such that the pair $( u_1, u_2 )$ lives on the Jacobian
of a genus two-curve. It will also be natural to impose
a constraint which amounts to looking at sub-varieties
of the Jacobian. At the end of this it will, nevertheless,
be possible to recover the sequence of zeroes and poles in 
(\ref{arrayzer}) and (\ref{arraypol}).

\subsection{ Geometry and Automorphisms
of Hyper-elliptic curve for the fuzzy $S^4$  }

 The equation (\ref{eqwrts}) defines a Riemann surface
 of genus $3$. The integrals of interest (\ref{integsig}) and (\ref{integt}),
 are integrals of a holomorphic differential along certain cycles of
the Riemann surface.
 It is useful to recall
the  Riemann-Hurwitz formula
\be
(2g-2) = n ( 2G -2 ) + B\;,
\label{riehur}
\ee
which gives the genus $g$ of the covering surface in terms
 of the genus $G$ of the target
and the number of branch points $B$.   Since the RHS
of  (\ref{eqwrts}) is a polynomial of degree $8$ there are
 $8$ points
 where $ s = 0 $, i.e. $8$ branch points. Here $G=0$ since the $r$
 co-ordinate can be viewed as living
on the sphere, $n = 2 $ and $B=8$. So the integrals are defined
 on a genus $g=3$ curve, which we will call $ \Sigma_3$.

After dividing out by $r_0^8$ on both sides of (\ref{eqwrts})
and rescaling $ s =  { r_0^4 \ts \over 1 + r_0^4 } $, $ r^4 =  r_0^4 \tr^4 $,
$ r_1^4 =   r_0^4 \tr_1^4 $ we have
\be
\tilde s^2 =  ( {\tilde r}^4 - 1 )( { \tilde r }^4 - { \tilde r_1}^4 )\;.
\label{rescaeq}
\ee
There are three independent holomorphic differentials
on this curve
$ \omega_1 = { d\tr \over  \ts }  , \omega_2 =  { \tr d \tr  \over \ts }  ,
 \omega_3 = { \tr^2 d\tr \over \ts }  $
(see for example \cite{farkra,baker}).
 The infinitesimal time elapsed
is  $ dt = { dr \over s }  = { ( 1  + r_0^4 ) \over r_0^3 }
{ d \tilde r \over  \tilde s } $.
It will be important in calculating the integrals  
(\ref{integt})(\ref{integsig}), to
understand the automorphisms of the Riemann surface $ \Sigma_3 $.
In the limit of large $r_0$, ${ \tr_1}^4 $  approaches $-1$ and
 there is an automorphism
\bea
&  \tr \rightarrow  { 1 \over \tr }  \nn\\
&  \ts \rightarrow { i \ts \over \tr^4 }  \nn \;,
\eea
which leaves the equation of the curve unchanged.
 It transforms the holomorphic differentials as follows
\begin{eqnarray}
 \omega_1 &\rightarrow& i \omega_3 \nonumber\\
 \omega_2 &\rightarrow& i \omega_2 \nonumber\\
 \omega_3 &\rightarrow& i \omega_1 \nonumber\;.
\end{eqnarray}
For any finite $r_0$ there is a $Z_2 $ automorphism,
$ { \tr } \rightarrow - { \tr }  $.
Quotienting by this can be achieved by changing variables
$ { \tr}^2 = x $.
The first and third holomorphic differentials $ \omega_1 $ and $\omega_3$
transform as follows
\bea
&&{ d{ \tr } \over \sqrt {  ( \tr^4-1 ) ( \tr^4 - \tro^4 )}}
 =  { dx \over \sqrt { 4 x ( x^2 -1 ) ( x^2 - \tro^4 ) } } \nn\\
&& { \tr^2  d{ \tr } \over \sqrt {  ( \tr^4-1 ) ( \tr^4 - \tro^4 )}}
 =  { x dx  \over \sqrt { 4 x ( x^2 -1 ) ( x^2 - \tro^4 ) } }\;.
\eea
We can view this in terms of a map from a genus three curve $( \tr , \ts)$
to a genus two  curve $ ( x , y )$
\bea
x &=& \tr^2 \nn\\
y &=& 2 \tr \ts\;,
\label{fstmap}
\eea
which implies
\be
y^2 = 4 \tr^2 ( \tr^4 - 1 ) ( \tr^4 - {\tro}^4 )
    = 4 x ( x^2 -1 ) ( x^2 - {\tro}^4 )\;.
\label{twdsgenpm}
\ee
This genus $2$ curve $ (x,y)$ will  be called $ \Sigma_2$.
The  holomorphic differentials are  related by
  $ { d \tr \over \ts }  = {dx  \over y }$ and $ { \tr^2 d\tr \over \ts } = { x dx \over y }$.
The map (\ref{fstmap})  has branching number
 $B=0 $, in agreement with (\ref{riehur}). Even though $ { \partial x \over \partial \tr } = 0 $
at $\tr=0$, this is not a branch point, since $x $ is not a good local
co-ordinate
for the curve $ ( x , y )$ at $x = 0 $. Rather, a good local  co-ordinate
is $ y$ which is linearly related to $\tr$ and this means that there is no
branch point.

The second holomorphic differential transforms as
\be
{ \tr  d \tr \over \sqrt { ( \tr^4 - 1 ) ( \tr^4 - r_1^4 )} }
 = { dx \over 2 \sqrt { ( x^2 -1 ) ( x^2 - \tro^4 ) } }\;.
\ee
This can be viewed in terms of a map
\bea
 \tilde x  &=& \tr^2 \nn\\
\tilde y  &=&  \ts
\eea
to a target torus $(\tilde{x},\tilde{y})$, obeying
\be
{ \tilde{y}}^2  = ( { \tilde x }^2 -1 ) ( { \tilde {x} }^2 -  \tro^4 )\;.
\label{gen1base}
\ee
This map has $B = 4 $, again in agreement with (\ref{riehur}). There are two branch points  at
$ ( \tr  , \ts ) = ( 0 , \pm {\tro}^2 ) $
corresponding to
$ ( \tilde x  , \tilde y   ) = ( 0 , \pm {\tro}^2 ) $.
Similarly we have two branch points at  $ \tr = \infty $.
The region near $\tr = \infty $ is best studied by defining variables
$ \hr =  { \tr }^{-1}  $ and   $ \hs = { \ts { \tr }^{-4} }$
which re-express (\ref{rescaeq})
 as $\hs^2 = ( 1 - \hr^4 ) ( 1 - \hr^4 {\tro}^4 )$.
Similarly we define $ \hat {x} = { \tilde x }^{-1} ,
\hat {y } =  \tilde { y } { \tilde {x}}^{-2}   $
which re-expresses (\ref{gen1base})  as
 $ \hat {y}^2 = ( 1 - \hat {x}^2 ) ( 1- \hat {x}^2 \tro^4 ) $.
Two  branch points corresponding to $\tr = \infty $
 are at  $ ( \hat {x} , \hat {y }   ) =   ( 0 ,  \pm 1 ) $.

The genus two curve $ \Sigma_2 $  itself can be related to genus one curves.
We first rewrite the $(x,y)$ equation (\ref{twdsgenpm}) as
\bea
   y^{2}  =  4x(x-1)(x+1)(x - iR_1^2 ) ( x + iR_1^2 )\;,
\label{gentwo}
\eea
with $R_{1}^{2}= - i \tr_1^2$ and real. This
is of the special form  \cite{enolrev,torisplit}
\be
   y^{2}=x(x-1)(x-\a)(x-\b)(x-\a\b)\;,
\label{redform}
\ee
with $\a=-1$ and $\b=iR_{1}^{2}$. Such curves have an automorphism $T_1$
\bea
 && T_1(x) \equiv X =  {\a \b \over x } = { - i R_1^2 \over x } \nn\\
 && T_1 (y) \equiv Y =  { (\a \b)^{3 \over 2 }y \over x^3 } =
{e^{ i \pi \over 4 }  R_1^3 y \over x^3 }\;,
\label{T1def}
\eea
which transforms the holomorphic differentials
\bea
 T_1 \left( { dx \over y } \right) &&= { i X dX \over Y \sqrt { \beta } }  \nn\\
 T_1 \left( { x dx \over y } \right) &&= { - i \sqrt { \beta }  dX \over Y }\;.
\label{holactT1}
\eea
The genus two curve  (\ref{gentwo}) can be put in the form
(\ref{redform}) in yet another way, by choosing
$ \a = -1 $, $\b = -i R_1^2 $. This gives another automorphism
$T_2$ which acts as
\bea
T_2 ( x )  =  {   i R_1^2 \over x } \nn\\
T_2 ( y )  =  {  e^{ 3 i \pi \over 4 } y R_1^3 \over x^3 }   \;.
\label{deft2}
\eea
With either choice of $ ( \alpha,\beta ) $
one is led to look for variables which are invariant
under the automorphism and as a result one describes the genus
$2$ curve as a covering of { \it two } genus one curves
$ ( \xi_{\pm} , \eta_{\pm} )$ \cite{enolrev,torisplit}
\be\label{enol}
      \eta_{\pm}^{2}=\xi_{\pm}(1-\xi_{\pm})(1-k_{\pm}^{2}\xi_{\pm})
\quad\textrm{with}\quad k_{\pm}^{2}=-\frac{(\sqrt{\a}\mp\sqrt{\b})^{2}}
{(1-\a)(1-\b)}\;.
\ee
The maps are given by
\bea
\xi_{\pm} &=& \frac{(1-\a)(1-\b)x}{(x-\a)(x-\b)} \nn \\
\eta_{\pm} &=&  - \sqrt { ( 1 - \a ) ( 1 - \b )  }
{ x \mp \sqrt{\a \b } \over  ( x - \a )^2 ( x - \b )^2 } y \;.
\label{themapsgen1}
\eea
This gives an isomorphism of the Jacobian of the genus 2 curve
in terms of a product of the Jacobians of the  genus $1$ curves
$ \Sigma_{\pm }$. For the case $\alpha = - 1 $
 we have $ K ( k_+ ) = K^{\prime} (k_- ) $.
 Since the complex structure is
$ \tau =  i { K^{\prime } ( k )  \over K ( k ) }$,
this means that the complex structures of $ \Sigma_{+}$
 and $ \Sigma_{-}$ are related by the $ SL(2,\mathbb{Z}) $ transformation
$ \tau \rightarrow -{  1 \over \tau } $. Hence $ \Sigma_{\pm}$
have isomorphic complex structures.

As an aside we describe the full group of automorphisms of $ \Sigma_2$.
It includes $ \sigma $, the hyperelliptic involution,
$ \sigma ( x ) = x , \sigma ( y ) = - y $ ;
$T_3$, which acts as $T_3(x ) = - x , T_3 ( y ) = -i y $,
and $T_4 $, which acts as $ T_4(x) = -x , T_4 ( y) = i y $.
Relations in this group of automorphisms are
\bea
T_1 T_2 &=& T_3 \nn\\
T_2 T_1 &=& T_4 \nn \\
T_3T_4 = T_4T_3 &=& 1 \nn \\
T_3^2 &=& \sigma \nn \\
T_3 \sigma &=& T_4\;.
\eea

In the limit $r_0 \rightarrow \infty $,
$R_1 =1 $ and the equation for the curve simplifies
\be
y^2 = x ( x^4 - 1 )\;.
\ee
As a result, the automorphism group is larger than at finite $r_0$.
The automorphism group is generated by $ U_1$ and $U_2$
which act as follows
\bea
U_1 (x ) &=& { 1 \over x } \nn \\
U_1(y)  &=& { e^{ i \pi \over 2 } y \over x^3 }  \nn \\
U_2 ( x ) &=&  e^{ i \pi \over 2 } x  \nn\\
U_2 ( y ) &=&   e^{ i \pi \over 4 }  y\;.
\eea
If we write $ U_1 (x) = X, U_1 ( y ) = Y $,
we have the following action on the holomorphic
 differentials
\bea
{ dx \over y } &=& - i { X dX \over y } \nn\\
{ xdx \over y } &=& -i {  dX \over Y } \;.
\label{holactU1}
\eea
The action of $U_2$ on the holomorphic differentials
is just
\bea
 { dx \over y } &&\rightarrow { e^{ -i \pi \over 4 } dx \over y } \nn\\
 { x dx \over y } &&\rightarrow   { e^{ -3 i \pi \over 4 } xdx \over y }\;.
\eea
The automorphism group includes, as usual, the hyperelliptic involution
acting as $ \sigma ( x ) = x , \sigma(y) = - y $. There is also
an element $U_3$ acting as $ U_3(x) = -x , U_3 ( y) = i y $.
There are relations $ U_1^2 = \sigma  , U_2^2 = U_3 $.
In the large $r_0$ limit, $ R_1 \rightarrow 1 $, the
formulae for $T_1,T_2$ from (\ref{T1def}),(\ref{deft2}) simplify and
they can be written in terms of $U_1,U_2$. Indeed we find that
\bea
T_1 &=& \sigma U_2 U_1 \nn\\
T_2 &=& \sigma U_2^{-1}U_1 \nn\;.
\eea

\subsection{ Evaluation of integrals }

The time integral (\ref{integt}) can be done in terms of Appell functions.
 The indefinite integral is
\be
  { ( 1 + r_0^4 ) r \over \sqrt { ( 2r_0^4 + r_0^8 )}  }
 F_1 \left(  { 1 \over 4 };  { 1 \over 2 } , { 1 \over 2} ;  { 5 \over 4 } ;  { r^4 \over r_0^4 }  ,
-{ r^4 \over { 2  + r_0^4 } } \right) \;.
\label{appel1}
\ee
A quick way to get this is by using the Integrator \cite{integrator} but we outline a derivation. Expanding the integrand of (\ref{integt})
\bea
&&\int dr (  r_0^4 - r^4 )^{-{1\over 2 }} ( r_0^4 + 2 + r^4 )^{-{1 \over 2}}
\nn \\
&& =  r_0^{-2} ( 2 + r_0^4 )^{-\frac{1}{2}}   \int dr
  \sum_{k,l = 0}^{\infty}
  { \Gamma( { 1 \over 2 }) \over \Gamma ( { 1 \over 2 } - k )
 \Gamma( k +1 )  }
      { \Gamma( { 1 \over 2 })\over \Gamma ( { 1 \over 2 } - l )
\Gamma ( l+1 )  }
  (-1)^k  \left( {r \over r_0 }\right)^{4k} \left( { r^4 \over 2 + r_0^4 }\right)^l
\nn \\
&&
 =  r_0^{-2} ( 2 + r_0^4 )^{-{1 \over 2 }}\;
 r \sum_{k,l = 0}^{\infty}
{ 1 \over k! l! }  { \Gamma( { 1 \over 2 } )\over \Gamma
    ( { 1 \over 2 } - k )  } { \Gamma( { 1 \over 2 } )\over
   \Gamma ( { 1 \over 2 } - l )} { 1 \over 4k + 4l +1 } (-1)^k
  \left( {r \over r_0 }\right)^{4k} \left( { r^4 \over 2 + r_0^4 }\right)^l \nn\;.
\eea
Now use the following facts about $\Gamma$ functions
\bea
&& { \Gamma ( { 1 \over 2 } ) \over \Gamma (  { 1 \over 2 } - k ) }
   = (-1)^k { \Gamma ( { 1 \over 2 } + k  ) \over \Gamma (  { 1 \over 2 }  ) }
\nn \\
&& { \Gamma ( { 1 \over 2 } ) \over \Gamma (  { 1 \over 2 } - l ) }
   = (-1)^l { \Gamma ( { 1 \over 2 } + l  ) \over \Gamma (  { 1 \over 2 }  ) }
\nn \\
&& { 1 \over 4 ( k + l +{1\over 4 } ) } = { \Gamma ( { 5 \over 4 } )
 \over \Gamma ( { 1 \over 4 } ) }   { \Gamma ( { 1 \over 4 } + k + l  )
 \over \Gamma ( { 5 \over 4 }  + k + l ) } \nn \;,
\eea
to recognise the series expansion of the Appell function \cite{functions}
\be
F_1 ( a ; b_1 , b_2 ; c ; z_1 , z_2 ) = \sum_{k=0}^{\infty} \sum_{l = 0 }^{\infty }
{ (a)_{k+l} (b_1)_{k} (b_2)_{l} z_1^k z_2^l \over (c)_{k+l} k! l! }  \;,
\ee
with arguments as given in (\ref{appel1}) and where we have also made use of the Pochhammer symbols $(a)_n=\frac{\G(a+n)}{\G(a)}$.

Using (\ref{appel1}),
 the time taken to collapse from the initial radius $r_0$ to
a smaller radius $r$ is
\be
t = { ( 1 + r_0^4 ) r_0 \over \sqrt { ( 2r_0^4 + r_0^8 )}  }
 F_1 \left(  { 1 \over 4 };  { 1 \over 2 } , { 1 \over 2} ;  { 5 \over 4 } ;
1  , -{ r_0^4 \over { 2  + r_0^4 } } \right)
 - { ( 1 + r_0^4 ) r \over \sqrt { ( 2r_0^4 + r_0^8 )}  }
 F_1 \left(  { 1 \over 4 };  { 1 \over 2 } , { 1 \over 2} ;  { 5 \over 4 } ;  { r^4 \over r_0^4 }  ,
-{ r^4 \over { 2  + r_0^4 } } \right) \;.
\label{timetor}
\ee

For the special values $r=r_0$,  $z_1 =1 $, $F_1$ of (\ref{appel1}) simplifies to
\begin{equation}
{ ( 1 + r_0^4 )  \over r_0 \sqrt { ( 2 + r_0^4 ) } }
~~ {}_2F_1 \left( { 1 \over 4 } , { 1 \over 2} ;  { 5 \over 4 }  ;  1 \right) ~~
 {}_2F_1 \left( { 1 \over 4}  , { 1 \over 2}  ;  { 3 \over 4 }  ;    -{  r_0^4 \over 2 + r_0^4 }  \right)
\label{uplim} \;.
\end{equation}

For $r = 0 $, $F_1=1$ and the indefinite integral (\ref{appel1}) evaluates to zero. Hence the time of collapse is
given by (\ref{uplim}) which can be simplified to
\be\label{toc}
T = { ( 1 + r_0^4 )  \over  r_0 \sqrt { ( 2  + r_0^4 )} } { \Gamma ( { 5 \over 4 } )
\Gamma ( {1 \over 2} )
 \over {\Gamma ( {3 \over 4 } ) } }
{}_2F_1 \left( { 1 \over 4}  , { 1 \over 2}  ;  { 3 \over 4 }  ;    { - r_0^4 \over 2 + r_0^4 }  \right)      \;.
\ee
For large $r_0$ the last argument of the hypergeometric function simplifies to
$ -1 $ and we get
\be
T = r_0
{ \Gamma ( { 5 \over 4 } ) \Gamma ( {1 \over 2} ) \over \Gamma ( {3 \over 4 } )}
{ 2^{-1/4} \Gamma ( { 3 \over 4 } ) \over { \Gamma ( { 5 \over 8}   )   \Gamma ( { 5 \over 8 } ) }}
= r_0 \sqrt { \pi } { \Gamma ( { 9 \over  8 } ) \over  \Gamma ( { 5 \over 8 }  ) }
\sim 1.1636... ~~  r_0\;.
\label{tlrgr0}
\ee

It is also of interest to compute the interval in $\sigma $ along the
D-string from the minimum size of the funnel cross-section to the
place where the funnel blows up. The indefinite integral (\ref{integsig})  gives by direct evaluation, just as above
\be
\frac{i\;(1+r_0^4)}{r_0^2}\frac{r}{\sqrt{2+r_0^4}}F_1\left(\frac{1}{4},\frac{1}{2},\frac{1}{2},\frac{5}{4};\frac{r^4}{r_0^4},-\frac{r^4}{2+r_0^4}\right)\;.
\label{sigtor}
\ee
The distance to blow-up is given by the difference of the last expression  evaluated at infinity and at $r_0$. We do not have an exact formula for  the
large $r$  asymptotics of the Appell function at finite $r_0$. We will thus be forced to take the large $r_0$ limit immediately. This will reduce (\ref{sigtor}) to
\be
i\; r_0 \;r \;{}_2F_1\left(\frac{1}{8},\frac{1}{2},\frac{9}{8};\frac{r^8}{r_0^8} \right)
\ee
and the distance to blow-up will be
\be
\frac{i\; r_0\; \Gamma(\frac{3}{8})\Gamma(\frac{9}{8})}{\sqrt{\p}(-1)^{1/8}}-\frac{i\; r_0\; \sqrt{\p}\Gamma(\frac{9}{8})}{\Gamma(\frac{5}{8})}\;.
\ee
The final result is
\be
\S=r_0 (\sqrt{2}-1)\sqrt{\p}\frac{\G(\frac{9}{8})}{\G(\frac{5}{8})}\sim 0.4819..  r_0\;.
\ee
 Hence the full period $4\S$ is $ 1.9276.. r_0 $.
The time of collapse is $ 1.1636...  r_0  $. The
space and time periods are no longer the same as was
the case for  fuzzy $S^2$, since there is a relative factor of
$ ( \sqrt {2} -1 )$.  

\section{Reduction of the $ g =3 $ curve and inversion
  of the hyper-elliptic integral for the fuzzy-$ S^{4} $}

Whereas we have formulae for the time elapsed $t$ in terms of
$r$  (\ref{timetor}) or the D1-co-ordinate $\sigma$ as a function of $r$
by using (\ref{sigtor}),  it is desirable to have the inverse  formulae
expressing $r$ as a function of $ \sigma $ and $t$. It will turn out
that, as in the case of the fuzzy $S^2$, it will be useful to define
a complex variable  $ u_1 = \sigma - i t $. Whereas the
$u$ variable in the case of fuzzy $S^2$ lives on a genus one
curve, here the story will involve higher genus curves
and will require the introduction of a second complex variable $u_2$.

 \par
 Let us  revisit the $\sigma$-integral in the case of the fuzzy-$S^{4}$. The integral we want to perform is
  \be
\nn      \sigma =  \int_{r_{0}}^{r} \frac{(1+r_{0}^{4})dr}{\sqrt{(r^{4}-r_0^{4})(r^{4}-r_{1}^{4})}}
   \ee
with $r_{1}^{4}=-(2+r_{0}^{4})$. If we make the re-scaling $\tilde{r}=\frac{r}{r_{0}}$ and $\tilde{r}_{1}=-
\frac{2+r_{0}^{4}}{r_{0}^{4}}$ we get
\be
   \frac{ \sigma r_{0}^{3}}{(1+r_{0}^{4})}=\int_{1}^{\tilde{r}}\frac{d\tilde{r}}{\sqrt{(\tilde{r}^{4}-1)(\tilde{r}^{4}-\tilde{r}_{1}^{4})}}\;.
\ee
The RHS is the integral over the holomorphic differential on $\S_3$ of section $3.4$. Using the reduction to $\S_2$ by making a change of variables $\tilde{r}^{2}=x$, we arrive at
\be\label{genus2}
   \frac{ 2  r_{0}^{3}\; \sigma}{(1+r_{0}^{4})}=\int_{1}^{x}\frac{dx}{\sqrt{x(x^{2}-1)(x^{2}-\tilde{r}_{1}^{4})}}\;.
\ee
Similar steps for the time-dependent fuzzy-sphere give
\be\label{genus2t}
-i\frac{2r_0^3\; t}{(1+r_0^4)}=\int_1^x \frac{dx}{\sqrt{x(x^2-1)(x^2-\tr_1^4)}}\;.
\ee
At this point it is useful to introduce a complex variable $u_1=\s-i t$. The inversion of the integrals
(\ref{genus2}), (\ref{genus2t}) is related to the Jacobi inversion problem
\bea\label{jacob2}
\nn \int_{x_{0}}^{x_{1}}\frac{dx}{y}+\int_{x_{0}}^{x_{2}}\frac{dx}{y}&=&
 u_{1}\\
 \int_{x_{0}}^{x_{1}}\frac{xdx}{y}+\int_{x_{0}}^{x_{2}}\frac{xdx}{y}&=&
 u_{2}\;,
\eea
where $x_{0}$ is any fixed point on the Riemann surface $\S_2$.
We will set $x_0=1$. By further fixing $x_{2}=1$ we recover the
integral of interest in the first line
\bea
\label{restjacob}
\nn \int_{1}^{x_{1}}\frac{dx}{y} &=&
 u_{1}\\
 \int_{1 }^{x_{1}}\frac{xdx}{y} &=&
 u_{2}\;,
\eea
This is a constrained Jacobi inversion problem which is related to a
sub-variety \footnote{  The geometry of such
subvarieties is   discussed extensively  in \cite{Gunn,griha}.
One result is that (\ref{restjacob}) defines a complex analytic 
homeomorphism from $ \Sigma_2$ to a complex analytic submanifold 
of $ J ( \Sigma_2 ) $.  } of the Jacobian of $ \S_2$, denoted as 
 $ J ( \Sigma_2 ) $. 
A naive  attempt to consider the inversion of the first equation 
of (\ref{restjacob}) in isolation runs into difficulties with infinitesimal 
periods as explained on page 238 of \cite{baker}.

 By switching to the variables $ \xi_{\pm} , \eta_{\pm} $ defined in
(\ref{themapsgen1}),  the system (\ref{jacob2})
can be reduced to the sum of simple elliptic integrals
\bea\label{xis}
\nn\int_{\xi_{0}}^{\xi_{1}}\frac{d\xi_{+}}{\eta_{+}}+\int_{\xi_{0}}^{\xi_{2}}\frac{d\xi_{+}}{\eta_{+}}&=& u_{+}\\
 \int_{\xi_{0}}^{\xi_{1}}\frac{d\xi_{-}}{\eta_{-}}+\int_{\xi_{0}}^{\xi_{2}}\frac{d\xi_{-}}{\eta_{-}}&=& u_{-}\;,
\eea
where $ \xi_1 = \xi_{\pm} (x_1)  $ and $ \xi_2 = \xi_{\pm} (x_2) $.
We have used
\be\label{us}
\frac{d\xi_{\pm}}{\eta_{\pm}}=\sqrt{(1-\a)(1-\b)}(x\pm\sqrt{\a\b})\frac{dx}{y}\quad\textrm{and}
\quad u_{\pm}=\sqrt{(1-\a)(1-\b)}(u_{2}\pm\sqrt{\a\b}\;u_{1})\;.
\ee
The first of the two integrals can be brought into the form
\be
\int_{1}^{\sqrt{\xi_{1}}}\frac{2dz}{\sqrt{(1-z^{2})(1-k_{+}^{2}z^{2})}}+\int_{1}^{\sqrt{\xi_{2}}}\frac{2dz}{\sqrt{(1-z^{2})(1-k_{+}^{2}z^{2})}}=u_{+}\;,
\ee
by the substitution $\xi=z^{2}$ and then split to give
\bea
\nn 2\int_{0}^{\sqrt{\xi_{1}}}\frac{dz}{\sqrt{(1-z^{2})(1-k_{+}^{2}z^{2})}}&+& 2\int_{0}^{\sqrt{\xi_{2}}}\frac{dz}{\sqrt{(1-z^{2})(1-k_{+}^{2}z^{2})}}\\
&-& 4
\int_{0}^{1}\frac{dz}{\sqrt{(1-z^{2})(1-k_{+}^{2}z^{2})}}=u_{+}\;,
\eea
 which are just
 \be
2Sn^{-1}(\sqrt{\xi_{1}},k_{+})+2Sn^{-1}(\sqrt{\xi_{2}},k_{+})-4K(k_{+})=u_{+}\;.
\ee
Then by using the addition formulae for $Sn^{-1}$ functions \cite{byrd},
we arrive at
\be
Sn^{-1}\left(\frac{\sqrt{\xi_{1}}\sqrt{(1-\xi_{2})(1-k_{+}^{2}\xi_{2})}+\sqrt{\xi_{2}}\sqrt{(1-\xi_{1})(1-k_{+}^{2}\xi_{1})}}{1-k_{+}^{2}\xi_{1}\xi_{2}},k_{+}
\right)=\frac{u_{+}}{2}+2 K(k_{+})\;.
 \ee
 Thus by setting $x_{2}=1$ we get $\xi_{2}=1$ and
 \be
\frac{1-\xi_{1}}{1-k_{+}^{2}\xi_{1}}=Sn^{2}\left(\frac{u_{+}}{2}+
2K(k_{+}),k_{+}\right)\;.
\label{solwthellip}
\ee
 After using the fact that $Sn$ is anti-periodic in
$2K(k)$ and implementing half-argument formulae, we end up with
\be\label{sol}
\frac{1+Cn(u_{+},k_{+})}{1+Dn(u_{+},k_{+})}=\xi_{1}\;.
 \ee
Starting from the second integral of (\ref{xis})
the same steps lead to
\be\label{sol2}
\frac{1+Cn(u_{-},k_{-})}{1+Dn(u_{-},k_{-})}=\xi_{1}\;.
\ee
 These expressions hold the answer to the Jacobi
inversion problem, although we still need to decouple $u_{1}$ and
$u_{2}$ in the $u_{\pm}$'s.

\subsection{Series expansion of $u_{2}$ as a function of $u_{1}$}
The equations (\ref{sol}) and (\ref{sol2}) imply that $u_+$ and $u_-$ are constrained by
 \be\label{con}
\xi_{1}=\frac{1+Cn(u_{+},k_{+})}{1+Dn(u_{+},k_{+})}=\frac{1+Cn(u_{-},k_{-})}{1+Dn(u_{-},k_{-})}\;.
\ee
This can be used to solve for $u_2 (u_1)$ (or $u_1 (u_2)$). We do not have an explicit general solution to this transcendental
 constraint but we can solve it in  a series
 around the initial radius
 $x=1$. This corresponds to small times, i.e. corresponds to doing perturbation theory for $u_{1}$ around zero
and similarly for $u_{2}$, as can be seen from the
 integrals (\ref{restjacob}). We find 
\bea\label{serexp}
    \nn u_{2}&=&u_{1}+\frac{1}{6}(1+R_{1}^{4})u_{1}^{3}+\frac{1}{120}(1+R_{1}^{4})(7+3R_{1}^{4})u_{1}^{5}\\
    &&+\frac{1}{2520}(1+R_{1}^{4})(65+66R_{1}^{4}+9R_{1}^{8})u_{1}^{7}+\mathcal{O}(u_{1}^{9})\;.
\eea
As a consistency check, we can invert the two integrals independently and see whether the expansion (\ref{serexp}) agrees with the results.
\par
In more detail we have
 \be
  \nn\int_{1}^{x_{1}}\frac{dx}{\sqrt{x(x^{2}-1)(x^{2}+R_{1}^{4})}}=u_{1}\;,
\ee
which can be expanded and inverted to give
 \bea\label{xu1}
 \nn  x_{1}&=&\frac{1}{2}(1+R_{1}^{4})u_{1}^{2}+\frac{1}{24}(1+R_{1}^{4})(7+3R_{1}^{4})u_{1}^{4}+\frac{1}{360}(1+R_{1}^{4})
(65+66R_{1}^{4}+9R_{1}^{8})u_{1}^{6}\\
&&+\frac{1}{4320}(1+R_{1}^{4})(4645+7479R_{1}^{4}+3087R_{1}^{8}+189R_{1}^{12})u_{1}^{8}+\mathcal{O}(u_{1}^{10})\;.
\eea
The same can be done for
\be \nn \int_{1}^{x_{1}}\frac{x
dx}{\sqrt{x(x^{2}-1)(x^{2}+R_{1}^{4})}}=u_{2}\;,
\ee
 giving
\bea\label{xu2}
 \nn  x_{1}&=&\frac{1}{2}(1+R_{1}^{4})u_{2}^{2}-\frac{1}{24}(1+R_{1}^{4})(R_{1}^{4}-3)u_{2}^{4}+\frac{1}{360}(1+R_{1}^{4})
(9+6R_{1}^{4}+5R_{1}^{8})u_{2}^{6}\\
&&+\frac{1}{4320}(1+R_{1}^{4})(189+63R_{1}^{4}-297R_{1}^{8}-235R_{1}^{12})u_{2}^{8}+\mathcal{O}(u_{2}^{10})\;.
\eea
The expressions (\ref{xu1}) and (\ref{xu2}) can be combined to give the desired expansion of $u_2 (u_1)$. We find perfect agreement between this and (\ref{serexp}).

\subsection{Evaluation of  the time of collapse and the distance to blow-up}
We can use the previous discussion to give a new calculation of the time of collapse and the distance to blow-up, which can be shown to satisfy some non-trivial checks against the formulae in section 3. Consider (\ref{con}) at the limits $x_{1}=1$ and $x_{1}=0$, where $\xi_{1}=1$ and $\xi_{1}=0$. These
lead to the two equations
 \be
Cn(u_{\pm},k_{\pm})=0\quad\textrm{and}\quad
Cn(u_{\pm},k_{\pm})=-1\;,
\ee
 which give
 \be\label{uplusminus1}
  u_{\pm}^{x_{1}=1}=   2(2
m_{\pm}+1)K(k_{\pm})+4in_{\pm}K'(k_{\pm})     \qquad\textrm{and}\qquad
u_{\pm}^{x_{1}=0}=0\;.
\ee

We should recall that for our case of $\a=-1$, it turns out that $k_{+}$ is the complementary modulus of $k_{-}$, i.e. $k_{+}^{2}+k_{-}^{2}=1$, which in turn means that
\be
K'(k_{\pm})=K(k_{\mp})\;,
\ee
which is very useful in simplifying (\ref{uplusminus1}). From (\ref{us}) we can write
\be
u_{1}=\frac{u_{+}-u_{-}}{2\sqrt{(1-\a)(1-\b)}\sqrt{\a\b}}\;.
\label{bascalt}
\ee
Then by collecting like terms, the time of collapse or the distance to blow-up,
 will be extracted from imaginary or real values of $u_1$
\bea
\nn u_{1}&=&u_{1}^{x_{1}=1}-u_{1}^{x_{1}=0}\\
         &=& \frac{(2m_{+}- 2in_{-}+1)K(k_{+})- (2m_{-}- 2in_{+}+1)K(k_{-})}{\sqrt{-2(1-\b)\b}}\;.
\eea
In order to compare this with something that we already know, we will consider the large-$r_{0}$ limit,
 where $\b=i$ and $k_{\pm}^{2}=\frac{1\mp\sqrt{2}}{2}$. For these values we find that
\be\label{above}
  K(k_{-})=(\sqrt{2}-i)K(k_{+})\;.
  \ee
\par
   Note that we should be careful here and observe
 a subtlety: By taking $r_0\rightarrow\infty$, we make
 $k_-$ \emph{exactly} real and larger than 1, while at very
large values of $r_0$, $k_-$ has a small but non-zero positive
imaginary part. It will be useful to recall the properties of
 the complete elliptic integral of the first kind $K(k_-)$,
with $k_-$ real. This function has a branch cut extending
 from $1$ to $\infty$. This means that the value above
 and below the cut will differ by a jump. 
 This statement translates to \cite{functions}
\bea
\nn \lim_{\e\to 0^+} K(k_- - i\e) &=&K(k_-)\\
\lim_{\e\to 0^+} K(k_- + i\e) &=&K(k_-)+2iK'(k_-)\;.
\eea
Note that we are using the conventions for $K(k_{\pm})$ 
of  \cite{byrd}\cite{grrhyz} which differ from those of
\cite{functions} by a $ k \rightarrow k^2 $. 
We can numerically verify that what we have
  corresponds to the second case and we will
 thus amend the above equation (\ref{above}) to
\be
  K(k_{-})=(2i+\sqrt{2}-i)K(k_{+})\;.
  \ee
 We are now ready to proceed. After some minor algebra and with the use of the $\tan{\frac{3\p}{8}}=1+\sqrt{2}$, we reach
\be\label{uone}
 u_{1}=-2^{-3/4}\cos{\frac{3\p}{8}}(A+iB)K(k_{+})\;,
\ee
 where $A=A_{1}+\sqrt{2}A_{2}$ and $B=B_{1}+\sqrt{2}B_{2}$
with
 \bea\label{AS}
\nn A_{1}=2m_{+}+2n_{-}+2m_{-}-6n_{+}+2\;, && A_{2}= 2n_- - 2n_+   \\
B_{1}= 2m_{+}-2n_{-}-6m_{-}-2n_{+}-2\;, && B_{2}=2m_{+}-2m_{-}\;.
\eea
Now recall from (\ref{genus2}) that since $t$ is real and for the
 large-$r_{0}$ limit $t=-r_{0}u_{1}/2i$, we need
$u_{1}$ to be purely imaginary in order to reproduce the results that
 we already have.
 This can be simply done by setting $A=0$ which is satisfied by
\bea\label{vana}
\nn  n_{+}&=& n_-\\
m_{+}+m_-&=& 2n_{+}-1
\eea
and has a first simple solution if we
 choose $m_{-}=-1$ and $m_{+}=n_{+}=n_{-}=0$.

Then (\ref{uone}) becomes
\be
u_{1} = -i\;2^{-3/4}\cos{\frac{3\p}{8}}(4+2\sqrt{2})K(k_{+})
\label{newcompcol}
\ee
and with the relation $ u_1 = \sigma - i t $
 the time of collapse for large-$r_{0}$ follows
\bea
\nn T&=&\frac{r_{0}}{2}K\left(\sqrt{\frac{1-\sqrt{2}}{2}}\right)
\cos{\frac{3\p}{8}}2^{-3/4}(4+2\sqrt{2})\\
&\simeq& 2.3272..\frac{r_0}{2} = 1.1636.. r_{0}\;,
\eea
exactly what we got previously from the evaluation of the integral in terms of Appell functions. When the conditions (\ref{vana})  for the vanishing of $A$ are satisfied,
the expressions for $B_1$ and $B_2$ simplify to
\bea
B_1 = -4(2m_- + 1)   \hskip.2in   B_2 = -2(2m_- + 2n_+ +1)\;.
\eea
Setting $n_+=0$ we have $ B_1 $ and $B_2$ both proportional to
$ ( 2m_- + 1 )$. This means that the first collapse to
zero is repeated  after every interval of twice the initial
collapse time. This behaviour was anticipated using continuity
and the symmetries (\ref{arrayzer}).

Similarly,  we can set $B=0$ and recover a real expression which
 will correspond to the distance that it takes for the funnel which has
a minimum cross-section of $r_0$
  to grow to infinity. We first use the large $r_0$ limit.
 The reality conditions are
\bea
\nn m_{+}&=&m_{-}\\
n_{+}+n_-&=&-(2m_{-} + 1)
\eea
 and a simple solution is the one with
$m_{-}=m_{+}=n_+=0 $ and $ n_{-}=-1$. Then
 \bea\label{space}
\nn\S&=&r_{0}K\left(\sqrt{\frac{1-\sqrt{2}}{2}}\right)\cos{\frac{3\p}{8}}2^{-3/4}\sqrt{2}\\
&\simeq& 0.9639..\frac{r_0}{2} =0.4819..r_{0}\;,
\eea
again in agreement with the previous results.
Generally when the conditions for the vanishing of the
imaginary part  are satisfied,  expressions for $A_1$ and $A_2$ simplify
\bea
A_1 = -8 n_+   \hskip.2in A_2 = -2(2m_+ +2n_+ +1) \nn
\eea

Setting $n_+=0$ gives a sequence of poles at distances
proportional to $2m_+ + 1$. This was  anticipated from symmetry and continuity
using (\ref{symmsqual}) in section 3 and is shown in Figure 2.
Note that the pattern of zeroes along  the time-axis is the same
as for the fuzzy $S^2$. The pattern of poles along the space axis
is also the same as for the fuzzy $S^2$. The difference is that the
time from maximum radius to zero in the time evolution
is not identical to the distance from minimum radius to infinite radius
as in the case of fuzzy $S^2$. There is, nevertheless, a simple ratio
of $ ( \sqrt{2} -1 ) $ at the large $r_0$ limit.
\par
It is also important to stress that we can use the above solutions to
 study the time of collapse for  finite $r_0$  and
 then compare with what one gets
 from (\ref{toc}). We have checked numerically, for many values of $r_0$ between zero and $\infty$,  that the time given in (\ref{toc}) agrees to six decimal digits with
the expression
\be
T = { (1+r_0^4) \over 2 r_0^3 } {  (  K (k_+) + K(k_-) ) \over
\sqrt { -2 \beta ( 1 - \beta  ) } }\;,
\ee
which reduces to (\ref{newcompcol}) at large $r_0$.
Similarly for the distance to blow-up we have
\be
\Sigma = { (1+r_0^4) \over 2 r_0^3 }
{( 2 i +1 ) K ( k_+ ) - K(k_- ) \over \sqrt{ -2 \beta ( 1 - \beta ) }  }\;.
\ee
 \begin{figure}[t]\label{fig:ells4}
\begin{center}
\includegraphics[height=9cm,width=16cm]{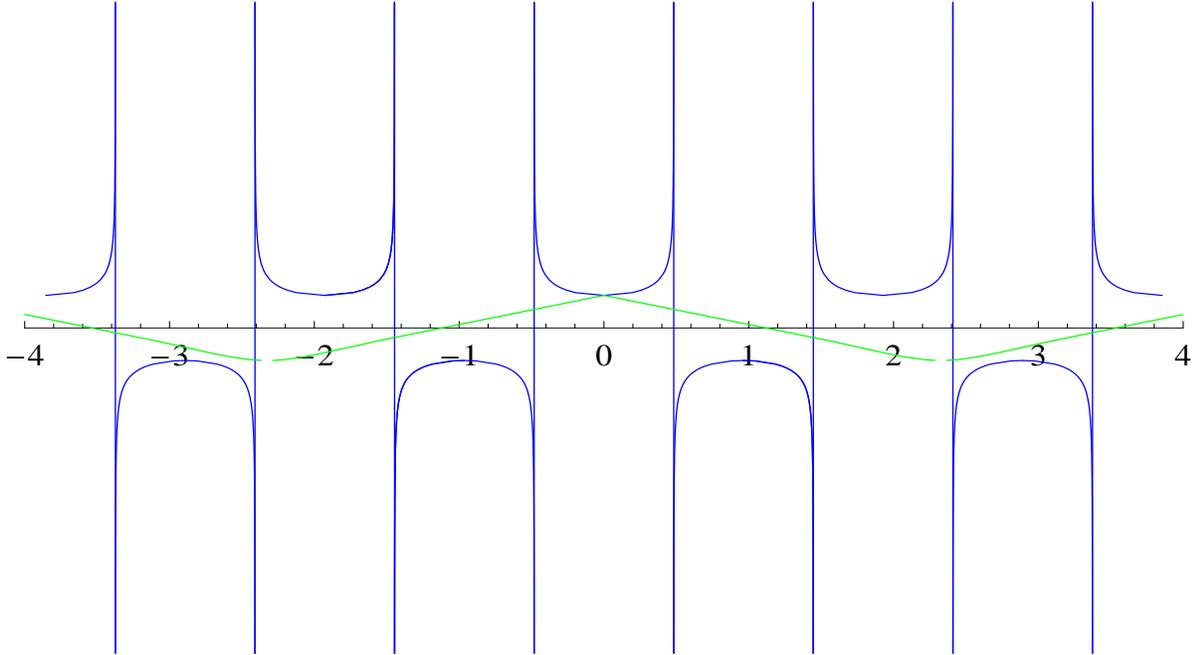}
\caption{Numerical plot of the Jacobi elliptic function solution for the static $S^4$-fuzzy funnel array and the collapsing $S^4$-sphere at large $r_0$.}
\end{center}
\end{figure}
We believe the same numerical agreement to hold for finite $r_0$. However, since the asymptotics of (\ref{sigtor}) for large $r$ at finite $r_0$ 
were not well available with our mathematical software, 
we cannot confirm this.
\par
So far we have considered the time of collapse where $r=0$
( or the distance to blow-up in the spatial case ), but we can also
consider $t$ as a function of $r$ for any finite $r$ in terms of the
Appell function (\ref{appel1}). The inverse expression
of $r$ in terms of $t$, and more generally the complex
variable $u_1 = \sigma - i t $, is contained in (\ref{solwthellip}) 
and the constraint (\ref{con}).

\subsection{Solution of the problem in terms of
 the $u_2$ variable and large-small duality }
We have seen that inverting the integrals (\ref{integt}) and (\ref{integsig})
requires the definition of a complex variable $u_1 = \sigma_1 - i t_1$ whose
real and imaginary parts are related to the space and time variables. 
In addition we have to introduce the second holomorphic differential, 
so that  there are $2$ complex variables $u_1, u_2 $. 
These variables 
are defined in terms of integrals and thus are subject to
identifications  by the period lattice $\cal L$ of  integrals
 around the $a$ and $b$-cycles 
of the genus $2$ Riemann surface, $\S_2$. They live on
  $ { \mathbb{C}}^2/{ \cal L } $, the Jacobian of $\S_2$.  Introduction of a second point $x_2$
on the Riemann surface relates our problem to the standard Jacobi inversion 
problem. The constraint  $x_2 =1 $ restricts to a subvariety of the Jacobian. 
So far the constraint (\ref{con}) has been viewed as determining $u_2 $ in terms of $u_1$.
This allowed us to describe $x_1$ as a function of $u_1$ in the neighbourhood
of $x_1 =1$ and also to get new formulae for the time of collapse/distance to blow-up
which have been checked numerically against evaluation of the integrals using Appell functions.
 However, it is also of interest to consider using the constraint to solve $u_1 $ in terms of $u_2$ and hence describe $x_1$ as a function of $u_2$. The reason for this is that the
 automorphisms of the Riemann surface allow us to relate the large $r$ (equivalently large $x_1$)
behaviour of the spatial problem described in terms of the $u_1$ variable, to
the small $r$ (equivalently small $x_1$) behaviour  of the time dependent problem described
 in terms of the $u_2$ variable and vice versa. The relation is simpler in the large $r_0$
 limit, hence  we describe this first and then we return to the case of finite $r_0$.
\par
The use of the $u_2$ variable is natural if we introduce another Lagrangian
 of the type (\ref{lagsfrhol}), with  $ \alpha =2 $. The
original Lagrangian of interest with $ \alpha = 0 $ and the
$ \alpha =2 $ Lagrangian are coupled in the Jacobi Inversion
problem (\ref{jacob2})
 as well as the constrained version of the Jacobi Inversion
problem  (\ref{restjacob}) obtained by setting $x_2 =1$.
 Just as $u_1$ is the
 complexified variable for the first Lagrangian, $u_2$ is the
complexified space-time variable for the second Lagrangian. 
It will be convenient, for the following discussion, to define 
the relation between the second set of space-time variables $ t_2 , \sigma_2 $ 
and the complex variable $u_2 $  as $ u_2 = -\sigma_2 + i t_2 $.
\par
We have also seen that
the automorphism $U_1$  of section 3.4 maps $du_1$ to
$-idu_2$ as in (\ref{holactU1}), hence  takes the time
dependent problem for the first kind of Lagrangian to the static
one for the second mapping zeroes of  a given
 periodicity to poles of the same periodicity.
 We will check this by investigating what happens to $u_2$ when we follow the collapse of $x_1$ down to
zero along imaginary $u_2$, or the blow-up of $x_1$ to infinity along real $u_2$. It is clear from the integrals in (\ref{restjacob}) (and the series expansion (\ref{serexp})) that when $x_1$ decreases from 1 to 0, both $u_1$ and $u_2$ are imaginary and when $x_1$ increases from 1 to infinity, both are real.
\par
Solving for $u_2$, the second of the equations	 (\ref{us}) will give
\be
   u_2= \frac{u_+ + u_-}{2\sqrt{(1-\a)(1-\b)}}
\ee
and by following the steps leading to (\ref{uone}) and (\ref{AS}) we arrive at
\be
   u_2= - 2^{-3/4}\cos{\frac{\p}{8}}(A'+i B')K(k_{+})\;.
\ee
Here we used $tan{\frac{\p}{8}}=\sqrt{2}-1$. The $A'$s and $B'$s are given by
\bea\label{APS}
\nn A'_{1}=2m_{+}+2n_{-}+2m_{-}-6n_{+}+2\;, && A'_{2}= 2 n_+ -2 n_-   \\
B'_{1}= 2+6 m_- +2n_- -2 m_+ +2 n_+\;, &&
B'_{2}=2m_{+}-2m_{-}\;.
\eea
 Notice by comparing (\ref{AS})  and (\ref{APS}),
 that  $A_1^{\prime} = A_1 , B_1^{\prime} = -B_1$ and
$A_2^{\prime}= -A_2 , B_2^{\prime} = B_2 $.
By choosing $m_-=-1$, $m_+=n_-=n_+=0$ we get
\bea
 \nn  u_2 &=& 2 i (2 - \sqrt{2} )
 \cos{\frac{\p}{8}}2^{-3/4}K\left(k_+\right)\\
 &\simeq& i\;0.9639..\;,
 \eea
which is \emph{exactly} the same as in (\ref{space})
 in agreement with what we expect from the automorphism. Similarly
 for the real case
\bea
 \nn  u_2 &=& - 2\sqrt{2}\cos{\frac{\p}{8}}2^{-3/4}K\left(k_+\right)\\
 &\simeq& -  2.3271..\;.
 \eea
 With the definition $ u_2 = - \sigma_2 + i t_2 $ we see 
 that we are getting the expected positive time of collapse 
 in terms of the alternative time-variable and the expected
 positive distance to blow-up with the alternative spatial variable.
The matching of the $A^{\prime}$ and $B^{\prime}$ 
 with $A$ and $B$ guarantees  
 that $x_1$, expressed as a function of $u_2$,
will have zeroes along the imaginary axis at odd integer multiples of 
a basic time of collapse and poles along the real axis at
 odd integer multiples of a basic distance to blow-up.
The time of collapse for $u_2$ is the same as the distance to blow-up
 for $u_1$ and the distance to blow-up for $u_2$ is the same
 as the time of collapse for $u_1$. This gives a precise map between 
 the behaviour at zero and the behaviour at infinity of the radius 
 of the fuzzy sphere, generalising the relations that were found for the case
 of the fuzzy $2$-sphere. This relation is expected from the 
automorphism $U_1$  for   large $r_0$
described in section 3,   which maps $ u_1 $ to $-iu_2$ (\ref{holactU1}).
This precise large-small relation for space and time-dependent fuzzy spheres 
is physically interesting. The physics of the large $r$ limit is 
very well understood because it corresponds to the D-strings blowing up
into a D5-brane. The physics of the small $r$ limit appears
mysterious because it involves sub-stringy distances. We have shown
that the two regions are closely related through the underlying Riemann
 surface which unifies the space and time aspects of the problem.

\subsubsection{Large-small duality at finite  $r_0$ }

 We saw in the discussion above that the spatial 
 problem of the fuzzy $S^4$ evolving from a minimum
 size at $r_0$ to infinity can be related to the time-dependent  
 problem  of the fuzzy sphere collapsing from $r_0 $ to zero. 
 This large-small relation involves a map between the $u_1$ description
 and the $u_2$ description of the problem and uses a simplification
 which is valid at large $r_0$. There continues to be a large-small 
 duality at finite $r_0$, but it is slightly more involved than the 
 one at large $r_0$. The difference is due to the
nature of the  automorphisms of the Riemann surface at finite $r_0$
and in the large $r_0$ limit, which were described in detail in section
3.4. Indeed, the discussion above used the automorphism $U_1$
in a crucial way.
\par
To describe the duality at finite $r_0$, it is useful to think of
a problem similar to the one we considered above, by choosing 
a different basepoint in (\ref{restjacob}), i.e. $x_0 = -iR_1^2 $
\bea
\label{restjacob2}
\nn \int_{-iR_1^2}^{  \tilde x_{1}}\frac{dx}{y} &=&
 \tilde u_{1}\\
 \int_{-iR_1^2}^{ \tilde x_{1}}\frac{xdx}{y} &=&
 \tilde u_{2}\;.
\eea
The upper limit is chosen, in the first instance, to vary  along the negative
imaginary axis up to zero. A second problem is to consider $\tilde x_1$
extending along the negative imaginary axis down to infinity.

Applying the automorphism $T_1$ of (\ref{T1def}) to  the first line of
(\ref{restjacob}) we have
\bea
u_1 (x_1) &=& \int_{1}^{x_1} { dx \over y } \nn\\
    &=& { e^{ i \pi \over 4 } \over R_1 }
         \int_{-iR_1^2}^{\tilde x_1 = -{ iR_1^2 \over x }}
                                { X dX \over Y } \nn \\
    &=& { e^{ i \pi \over 4 } \over R_1 }
                  \tilde u_2 \left( \tilde x_1 = -{ iR_1^2 \over x } \right) \nn\;.
\label{dualfin1}
\eea
Similarly we have
\be
  u_2(x_1) = R_1 e^{\frac{i \p}{4}} \tilde u_1 \left( \tilde x_1 = -{ iR_1^2 \over x } \right)\;.
\label{dualfin2}
\ee 
As we saw earlier in this section,
 the   solution to (\ref{restjacob}) can be described
 in terms of either the $u_1$ or the $u_2$ variable. 
Likewise the inversion of (\ref{restjacob2}) can be expressed
 in terms of either $ \tilde u_1$ or $\tilde u_2$. 
The action of the automorphism $T_1$ described 
above implies that the solution of  the spatial problem 
 (\ref{restjacob})   where $x_1$ evolves from $1$ to infinity 
along the real axis, when given in terms of the real part of the
 $u_1$ variable,
 maps to the evolution of
$ \tilde x_1$ from $ -iR_1^2 $ along the imaginary
axis to zero, as described by the   $\tilde u_2$ variable.
Similarly the time-dependent problem of $x_1$ evolving
from $1$ to zero when described in terms of the imaginary
part of the $u_1$ variable,
maps to the evolution of $ \tilde x_1$ along the imaginary axis
from $ -iR_1^2 $ to infinity, as described by the $\tilde {u}_2 $ variable.
This shows that there continues to be a large-small 
duality at finite $r_0$, but it relates the original problem 
with real $x$ to a problem with imaginary $x$.

\section{ Space and Time dependent Fuzzy-$S^{6}$ }
Here we will briefly talk about the nature of the solution in the case of the $D1\perp D7$ intersection, discussed by \cite{cook}, which involves the fuzzy-$S^6$.
Starting from the D-string theory point of view, we will now have
seven transverse scalar fields,
given by the ansatz
\begin{equation}\label{d7ansatz}
    \Phi^{i}(\s,\t)=\pm \hat{R}(\s,\t)G^{i}, \qquad i=1,\ldots
    7\;,
\end{equation}
where the $G^{i}$'s are given by the action of the $SO(7)$ $\Gamma^i$'s on the symmetric and traceless $n$-fold tensor product of the basic spinor $V$, the dimension
of which is related to $n$ by \cite{sphdiv}
\begin{equation}\label{Nn}
    N=\frac{(n+1)(n+2)(n+3)^2 (n+4)(n+5)}{360}\;.
\end{equation}
Again, the radial profile and the fuzzy-$S^{6}$ physical radius are related by
\be
 R_{ph}(\s ,\t)=\sqrt{c}\lambda\hat{R}(\s,\t)\;,
 \ee
 with $c$ the quadratic Casimir $G^{i}G^{i}=c\mathbf{1}_{N}=n(n+6)\mathbf{1}_{N\times N}$. The time-dependent generalisation for the leading $1/N$ action of \cite{cook} can be written in dimensionless variables, which are once more defined as in (\ref{defdimless}),
\begin{equation}\label{d17}
    S_{1}=-N\;T_{1}\int d^{2}\s
    \sqrt{1+r^{\prime 2}-\dot{r}^{2}}\left(1+r^{4}\right)^{3/2}\;.
\end{equation}
 In a manner identical to the discussion in section 3,
 the equations of motion can be given in a Lorentz-invariant expression
  \be
\pd_{\m}\pd^{\m}r+(\pd_{\m}\pd^{\m}r)\;(\pd_{\n}r)\;(\pd^{\n}r)-(\pd_{\m}
\pd^{\n}r)\;(\pd_{\n}r)\;(\pd^{\m}r)=6\;r^{3}\left(\frac{1+(\pd_{\m}r)\;
(\pd^{\m}r)} {1+r^{4}}\right)\;.
 \ee
 \par
At this point it is natural to propose that the form of the action and the equations of motion will generalise in a nice way for any fuzzy-$S^{2k}$ sphere
\be
S_{1}=-T_1\int d\s^2 STr \sqrt{\left(1+r'^2-\dot{r}^2\right) \left(1+r^4\right)^{k}}\;,
\ee
with the large-$N$ equations of motion

\be
\pd_{\m}\pd^{\m}r+(\pd_{\m}\pd^{\m}r)\;(\pd_{\n}r)\;(\pd^{\n}r)-(\pd_{\m}
\pd^{\n}r)\;(\pd_{\n}r)\;(\pd^{\m}r)=2k\;r^{3}\left(\frac{1+(\pd_{\m}r)\;
(\pd^{\m}r)} {1+r^{4}}\right)\;.
 \ee
\par
As we saw in the previous cases, there will be a curve related to the blow-up of the funnel, derived by the conservation of pressure if we restrict to static configurations and also to the corresponding collapse of a $D6$-brane by conservation of energy if we completely drop the space variable.
 We find that the curve determining the solutions is
\begin{equation}
s^2 =  ( r^4 - r_0^4 ) ( r^4 - r_1^4 ) (r^4 - r_2^4 )\;,
\label{s6eq}
\end{equation}
which is of genus 5 and where a factor of $(1+r_0^4)$ has been absorbed in the definition of $s$. The roots are given by
\bea
 ( 1 + r_0^4 ) &=& u_0^4 \nn\\
  ( 1 + r_1^4) &=& u_0^4 \eta \nn\\
 ( 1 + r_2^4 )&=& u_0^4 \eta^2 \;,
\eea
with $ \eta = \exp{ 2 i \pi \over 3 } $.

\begin{itemize}
\item{\bf Automorphism at large $r_0$ : }
\newline
At large $r_0$, we have $r_1^4 = r_0^4 \eta $
and $r_2^4 = r_0^4 \eta^2 $. Then there exists an automorphism
\bea
 R &=& { r_0^2 \over r }  \nn\\
  S^2 &=&  - { s^2  r_0^{12} \over r^{12} }  \nn \;.
\eea

It is convenient to define $ { \tilde r } = { r \over r_0 } $,
and $ { \tilde s}^2 = s^2 r_0^{12} $. In these variables
\bea
\nn  \tr \rightarrow { 1 \over \tr } \\
\nn   \tilde{s}\rightarrow {i\tilde{s}\over \tr^{6}}
\eea
and the action on the holomorphic differentials is
\bea
\nn \omega_1 &\rightarrow& i\omega_5\\
\nn \omega_2 &\rightarrow& i\omega_4\\
\nn \omega_3 &\rightarrow& i\omega_3\\
\nn \omega_4 &\rightarrow& i\omega_2\\
\nn \omega_5 &\rightarrow& i\omega_1 \;.
\eea

\item{ \bf Automorphism at $r_0=0 $ :}
\newline
Now we have
 \be
  s^2 = r^4 ( r^4 - r_1^4 ) ( r^4 - r_2^4 )
  \ee
and there is an automorphism
\bea
 R  &=& { r_1 r_2 \over r }  \nn\\
 S^2  &=& { s^2  (r_1 r_2)^8  \over r^{16} } \nn\;.
\eea
In this limit $ r_1^4 = ( \eta -1 ) $ and $ r_2^4 = ( \eta^2 -1 )$,
so in the formulae above we can write
$ r_1^4 r_2^4 = 3 $.

\item{\bf  Symmetry at finite $r_0$ : }
\newline
It is useful to write the curve in terms
of a variable $u$ defined by $ u^4 = ( 1 + r^4 )$
and to write $ u_0^4 = 1 + r_0^4 $.
Then we have
\be
s^2 =   ( u^4 - u_0^4 ) ( u^4 - u_0^4 \eta  )( u^4 - u_0^4 \eta^2  )\;.
\ee
A symmetry is $ v = { u_0^2 \over u } $,
$ {\tilde s}^2 = -{ s^2 u_0^{12}   \over u^{12} }$.
Expressing the symmetry in the $r,s$ variables, we
have $ R , S $ obeying the same equation (\ref{s6eq})
with
\bea
( 1 + R^4 ) &=& { (1 + r_0^4 )^2 \over ( 1 + r^4)  } \nn\\
S^2  &=& - { s^2 ( 1 + r_0^4 )^3 \over ( 1 + r^4 )^3}  \nn\;.
\eea
This reduces to the $ R = { r_0^2 \over r } $ for $r>>1$, $R>>1$, $r_0>>1$. 
 Unfortunately $ R $ is not a rational function of $r$,
 but an algebraic function
of $r$ involving fourth roots, hence it is not a holomorphic or 
 meromorphic function.
Hence  it is not possible to use this symmetry to map the
holomorphic differentials of the genus $5$ curve to those on genus $1$
 curves. We can still
make the change of variables $ x = r^2 $ to get a reduction down to genus
$3$, but we have not been able to reduce this any further.
\end{itemize}

Since we cannot reduce the curve down to a product of genus one curves,
we cannot relate the problem of  inverting the hyper-elliptic integral
to elliptic functions. We can nevertheless relate it to
the Jacobi inversion problem at genus $3$. We consider variables
$u_1, u_2 , u_3 $ defined as

\bea\label{jacobi}
\nn \int_{1 }^{x_{1}}\omega_{1}+\int_{1 }^{x_{2}}\omega_{1} +\int_{1 }^{x_{3}}
\omega_{1}&\equiv& u_{1} \\
\nn       \int_{1 }^{x_{1}}\omega_{2}+\int_{1 }^{x_{2}}\omega_{2} +\int_{1 }^{x_{3}}
\omega_{2}&\equiv& u_{2}\\
\nn
\int_{1 }^{x_{1}}\omega_{3}+\int_{1 }^{x_{2}}\omega_{3} +\int_{1 }^{x_{3}}
\omega_{3}&\equiv& u_{3}\;.
\eea

The variables $u_1, u_2 , u_3 $ live on the Jacobian of the genus three
curve, which is a complex torus of the form $ { \mathbb{C}}^3/{ \cal L } $.
The integrands appearing above live naturally on a Riemann surface which is a cover
of the complex plane, branched at 8 points. The lattice $\cal L$ arises from doing the integrals
around the  $a$ and $b$-cycles
of the Riemann surface. The equations (\ref{jacobi}) can be inverted
to express

\bea\label{jacsol}
x_1 &=& x_1 ( u_1 , u_2 , u_3 ) \nn \\
x_2 &=& x_2 ( u_1 , u_2 , u_3 )  \nn \\
x_3 &=&  x_3 ( u_1 , u_2 , u_3 )  \nn\;,
\eea
where $x_i ( u_1, u_2 , u_3 )$ can be given in terms of genus three
theta functions, or equivalently in terms of hyper-elliptic Kleinian functions \cite{enol}.
The system (\ref{jacobi}) simplifies if we set $x_2= x_3 =1 $.
These simplified equations define a sub-variety of the Jacobian
which is isomorphic to the genus $3$ Riemann surface we started with
\cite{Gunn}. The constraints $ x_2 = x_3 =1$ can be used to solve,
 at least locally near $x_1 =1 $, for $u_2 , u_3 $ in terms of $u_1$.
Then we can write  $x_1 ( u_1, u_2 , u_3 ) $ as $ x_1 (u_1)$.
This program was carried out explicitly in section 3, where
the higher genus theta functions degenerated into expressions in terms
of ordinary elliptic functions thanks to the reduction of the
genus three curve we started with,  to a product of genus one curves.
For completeness, we give a short review of the solution to 
the Jacobi Inversion problem in terms of higher genus 
$\vth $ functions and the related Kleinian $ \sigma$-function. 

\subsection{ Jacobi inversion problem and $\vth $ functions }
The general Jacobi inversion problem can be formulated as follows
 \cite{ baker, enolrev,enol}. For any hyper-elliptic
curve $\S$ of genus $g$, realised as a 2-sheeted cover over a Riemann sphere
\be
y^{2}=\sum_{i=0}^{2g+2}\lambda_i x^i=\prod_{i=1}^{g+1}(x-a_i)\prod_{i=1}^{g+1}(x-c_i)\;,
 \ee
with $a$'s and $c$'s being the branch points, between which we stretch the cuts, the system of integral equations
 \bea\label{jacob}
\nn \int_{a_{1}}^{x_{1}}\omega_{1}+\ldots+\int_{a_{g}}^{x_{g}}
\omega_{1}&\equiv& u_{1} \\
\nn\vdots & & \vdots\\
\nn\int_{a_{1}}^{x_{1}}\omega_{g}+\ldots+\int_{a_{g}}^{x_{g}}
\omega_{g}&\equiv & u_{g}\;,
 \eea
describes the \emph{invertible}
Abel map, $\mathit{U}:\S^g/S_g\longrightarrow\textrm{Jac}(\S)$, taking $g$ symmetric points from
the Riemann surface to the Jacobian of $\S$. The latter is just $\mathbb{C}^g/\mathcal{L}$, where
$\mathcal{L}=2\z\oplus 2\z'$ is the lattice that is generated by the non-degenerate periods of the holomorphic
 differentials, or differentials of the first kind, defined on the surface
\be
   2\z=\oint_{A_k} \omega_l\quad\textrm{and}\quad 2\z'={\oint_{B_k}\omega_l}\;,
\ee
with
\be
\omega_{i}=\frac{x^{i-1}dx}{y}\;, \qquad i=1,\ldots,g\;.
\ee
The period matrix is the $g\times g$ matrix given by $\t=\z^{-1}\z'$ and belongs to the
Siegel upper halfspace of degree $g$, having positive imaginary part and being symmetric. There is also a set of
canonical meromorphic differentials, or differentials of the second kind, naturally defined on the
Riemann surface
\be
  \xi_j=\sum_{k=j}^{2g+1-j}(k+1-j)\lambda_{k+1-j}\frac{x^k dx}{4y}\;,\qquad j= 1,\ldots,g\;,
\ee
 the periods of which are
\be
2\eta=-\oint_{A_k} \xi_l\quad\textrm{and}\quad 2\eta'=-\oint_{B_k}\xi_l\;.
\ee
Now consider $\textbf{\emph{m}},\textbf{\emph{m}}'\in\mathbb{Z}^g$ two arbitrary vectors and define the periods
\be
  \Omega (\textbf{\emph{m}} , {\textbf{\emph{m}}}^{' } )= 2 \z \textbf{\emph{m}}+ 2
 \z'\textbf{\emph{m}}' \quad \textrm{and}\quad E(\textbf{\emph{m}},
 {\textbf{\emph{m}}}'  )=2 \eta\textbf{\emph{m}}+2
\eta'\textbf{\emph{m}}'\;.
\ee
We can define the fundamental Kleinian $\s$-function in terms of higher genus Riemann $\vth$-functions
\be
\s(\textbf{\emph{u}})=Ce^{\emph{\textbf{u}}^T \k \emph{\textbf{u}}}\vth((2\omega^{-1})\emph{\textbf{u}}-\textbf{K}_a|\t)\;,
\ee
where $C$ is a constant, $\k=(2\z)^{-1}\eta$ and $\textbf{K}_a$ is the vector of Riemann constants with base point $a$, given by the Riemann vanishing theorem
\be
\textbf{K}_a=\sum_{k=1}^g\int_{a}^{a_i}d\textbf{v}\;,
\ee
with $d\textbf{v}=(2\z^{-1})( \textbf{\emph{$\omega_1$}},\ldots,\textbf{\emph{$\omega_g$}})^T$ being the set of \emph{normalised} canonical holomorphic differentials. Furthermore, the genus-g Riemann $\vth$-function with half-integer characteristics $[\varepsilon]=\left[\begin{array}{c} \varepsilon' \\ \varepsilon \\ \end{array} \right]=\left[ \begin{array}{ccc} \varepsilon_1 '& \ldots & \varepsilon_g ' \\ \varepsilon_1 & \ldots & \varepsilon_g \end{array}\right]\in \mathbb{C}^{2g}$ is defined as
\be
\vth[\varepsilon]  (\textbf{\emph{v}} | \tau )  = \sum_{\textbf{\emph{m}}\in\mathbb{Z}^g}\exp\;\p i \left\{ (\textbf{\emph{m}}+\varepsilon')^T
   \t \;(\textbf{\emph{m}}+\varepsilon')+2(\textbf{\emph{v}}+\varepsilon)^T (\textbf{\emph{m}}+\varepsilon')\right\}\;.
\ee
\par
The general
pre-image of the Abel map can be given  in a simple algebraic form as the
roots of a polynomial  in $x$, given by

\be\label{inv}
  \mathbf{P}(x;\textbf{\emph{u}})=x^{g}-x^{g-1}\wp_{g,g}
(\textbf{\emph{u}})-x^{g-2}\wp_{g,g-1}(\textbf{\emph{u}})-
\ldots-\wp_{g,1}
  (\textbf{\emph{u}})\;.
\ee
Here $\textbf{\emph{u}}=(u_{1},\ldots,u_{g})^{T}$  and the $\wp$'s are higher genus versions of the standard Weierstrass
elliptic $\wp$-functions,
which are defined as the logarithmic derivatives of the fundamental hyper-elliptic $\s$-functions
\be
\wp_{ij}(\textbf{\emph{u}})=-\frac{\partial^2 \ln \s(\textbf{\emph{u}})}{\partial u_i\partial u_j}\;,\quad\wp_{ijk}(\textbf{\emph{u}})=-\frac{\partial^3
 \ln \s(\textbf{\emph{u}})}{\partial u_i\partial u_j\partial u_k}\;,\ldots\;,\quad i,j,k,\ldots\;= 1,\ldots,g
\ee
and have the nice periodicity properties
\be
   \wp_{ij}(\textbf{\emph{u}}+\Omega({\textbf{\emph{m}},\emph{\textbf{m}}}'))=\wp_{ij}(\textbf{\emph{u}})\;,\quad
   i,j=1,\ldots,g\;.
\ee
We can make use of all this technology and give at least an implicit solution for the $S^6$ at the stage where we
have been able to reduce the problem from one of genus-5,  to one of genus-3. By equating the general polynomial with roots $x_1,x_2,x_3$ to the
polynomial with $\wp$-coefficients, we get
\bea
    \nn x_1+x_2+x_3=\wp_{33}(u_1,u_2,u_3)\\
    \nn x_1 x_2 x_3 =\wp_{23}(u_1,u_2,u_3)\\
    x_1 x_2+x_2 x_3+x_3 x_1=-\wp_{13}(u_1,u_2,u_3)\;.
\eea
As we have already mentioned, we can fix two of the three points to be $x_2=x_3=1$, to get
\bea
\nn x_1+2=\wp_{33}(u_1 ,u_2 ,u_3 )\\
\nn x_1 =\wp_{23}(u_1 ,u_2 ,u_3 )\\
2 x_1+2 =-\wp_{13}(u_1 ,u_2 ,u_3 )\;.
\eea
This implies two transcendental constraints which can be used to get a solution for $x_1$, in terms of the $\wp$'s as functions of $u_1$.

\section{ Summary and Outlook }

The space and time dependence of fuzzy spheres
$S^2$, $S^4$ and $S^6$ are governed by equations
which follow from the DBI action of D-branes.
These fuzzy spheres can arise as bound states of
D0 with D2, D4 or D6 on the one hand, or as
cross-sections of fuzzy funnels formed by D1 expanding into
D3, D5 or D7 on the other. The purely time-dependent process
has the simplest realisation as the collapsing-expanding $D0-D2p$-brane,
although it also arises as a time dependent process for the
$D1-D(2p+1)$ equations with no variation along the spatial $D1$-direction.
The first order equations of motion are closely related 
to some Riemann surfaces
and the infinitesimal time elapsed or the infinitesimal
distance along the D1, are given by a holomorphic differential
 on these surfaces.
We have given descriptions of the solutions in terms of
 elliptic functions or their higher genus generalisations.
We  showed that the space and time processes
 exhibit some very interesting large-small dualities closely
related to the geometry of the Riemann surfaces.
\par
In the case of $S^2$, the Riemann surface has genus one
and the duality is related to properties of Jacobi
elliptic functions known as ``complex multiplication
formulae'' \cite{Chandrasekharan}.
The genus $1$ Riemann surface that arises here has automorphisms,
holomorphic maps to itself, which are responsible for these properties.
We observed that the large-small duality is also directly related to
a transformation property of the second order differential equations.

In the case of $S^4$ we derived formulae for the distance $ \sigma(r) $
along the D1-brane as a function of the radius of the funnel's fuzzy sphere
cross-section
and the time elapsed  $t (r) $ as a function of the radius. These formulae
were expressed in terms of special functions such as Appell
and hypergeometric functions. It was useful to combine space and time
into a complex variable $u_1 = \sigma - i t $ which was related to a genus three Riemann surface, having a number of automorphisms.
These automorphisms were used to eventually relate the problem  to a product
of two genus one Riemann surfaces. After introducing a second complex variable
$u_2$,  the problem of inverting the integrals
to get a formula for the radius as a function of the complexified space-time
variable $u_1$ was related to a classical problem in Riemann surfaces,
 the Jacobi Inversion problem. This  can be solved, in general, using higher
 genus theta functions. The relation to a product of genus
one surfaces in the case at hand,  means that the
 inversion problem can be expressed in terms of the standard
elliptic integrals. This allowed us to give a
 solution of the inversion problem in terms of standard
 Jacobi elliptic functions. The solution involves
 a  constraint in terms of elliptic functions, which can be used to solve
 $u_2$ in terms $u_1$, or vice versa. This approach yields
a new construction of $ r ( u_1)$ as a series around $r = r_0$,
which agrees with a direct series inversion of the Appell function.
It also gives new formulae for the time of collapse  and distance to blow-up
in terms of sums of complete elliptic integrals, which  were 
checked numerically to agree with
 the formulae in terms of hypergeometric functions.
\par
The automorphism
which allows a reduction of the problem to one involving genus one surfaces
also relates   the large $r$ behaviour of the spatial (time-dependent)
solutions to the small $r$ behaviour of the time-dependent (spatial)
 problem. The introduction of the extra variable $u_2$,
 required to make a connection to the Jacobi
 inversion problem, also enters the description of this large-small duality.
 \par
Our discussion in the $S^6$ case was less complete, but a lot of the structure
uncovered above continues to apply. The integrals giving
$t(r)$ or $ \sigma(r ) $ are integrals of a holomorphic differential
on a genus $5$ Riemann surface. A simple $ R = r^2 $ transformation
maps it to a holomorphic differential on a genus $3$ curve.
The inversion problem of expressing $r$ in terms of
$ u_1 = \sigma -i t $ can be related to the Jacobi inversion problem
and a solution in terms of higher genus theta functions is outlined.
We did not find any automorphisms of the genus $3$
Riemann surface which would relate the problem to  one involving
holomorphic differentials on a genus $1$ curve. There do continue to exist
symmetry transformations of the curve but they are no longer
 holomorphic in terms of the $(r,s)$ variables.
\par
It will be interesting to  consider the full CFT description of these
funnels. Some steps towards the full CFT description of
these systems has  been made \cite{larus}.
There should be a boundary state describing the spatial
configuration of a D1-branes forming a funnel which blows up into a
D3-brane. One could start with the CFT of the multiple D1-branes
and consider a boundary  perturbation describing the funnel which opens into
a D3. Alternatively we could start with a CFT for the D3 and
introduce boundary perturbations corresponding to the
magnetic field strength and the transverse scalar excited on the
 brane, which describe the D1-spike. In either case the boundary
perturbation will involve the elliptic functions which appeared
in section $2$. Similarly the boundary perturbation for the
fuzzy $S^4$ case would involve elliptic functions associated
 with a pair of genus $1$ Riemann surfaces, of the type
described in section 4. One such boundary state corresponds to the
purely time-dependent solution and another to the purely
space-dependent solution. If the large-small duality  of the
DBI action  continues to hold in the CFT context,
this would give a remarkable relation between the zero radius limit
of  time-dependent
 brane collapse and the well-understood blow-up of
 a D1-brane funnel into a higher dimensional brane.
Analytic continuations from Minkowski to Euclidean space
 in the context of boundary states 
have proved useful in studies of the rolling tachyon \cite{Sen}. 
 We expect similar applications of the analytic structures described here. 

 \par
A complementary way to approach these dualities is to consider
the supergravity description. The time dependent system of
a collapsing D0-D2-brane, for example, should correspond to a time-dependent
supergravity background, when the backreaction of spacetime
to the stress tensor of the D0-D2 system is taken into account.
A Wick rotation of this background can convert the time-like co-ordinate
$t$ to a space-like co-ordinate. Similarly the spatial funnel
of a D1-D3 system has a supergravity description with
an $S^2$ having a  size that depends on the distance along the
direction of the D-string. The Wick rotated D2-background should
be related by a large-small duality to the D1-D3 system. This is a
sort of generalisation of $S^1$ T-duality to $S^2$.
Similarly, the $D0-D4$ system is related to the $D1\perp D5$.
 The  Jacobi Inversion problem which underlies our solutions
 has already arisen in the context of general relativity \cite{grqc}.
 It will be interesting to see if this type of  GR application
 of the Jacobi Inversion also arises in the spacetime solutions
 corresponding to our brane configurations. Another perspective 
on the geometry of these configurations , including the 
boosted ones in the Appendix A.1,  and on the relations 
to $S$-branes,  should be  provided by 
considering the induced metric on the brane as for example 
in \cite{wang}.

\par
Generalisation of this discussion to funnels of $\mathbb{CP}^2$ and other
cross-sections, as well as to funnels made of dyonic strings  will
be interesting.  The study of finite $n$ corrections as in \cite{rst}
is another interesting avenue.
\par
 Studying fluctuations around the solutions considered is likely 
to  involve equations where  the elliptic functions and
 their higher genus generalisations appear as potentials
 in Dirac or Klein-Gordon type equations for the fluctuations.
 Schr\"odinger equations with such potentials are well
 studied in the literature of integrable models \cite{dubrovin}.
 It will be interesting to see if such integrable models
 have a role  in the analysis of fluctuations.
\par
 Finally, the genus $3$ curve for the $S^4$ was very special.
 It is a related by the $ R =r^2$ map to a product
 of genus $1$ and a genus $2$ curve. The resulting
 genus $2$ curve itself has a Jacobian which
 is related to $K3$ Kummer surfaces and factorises
 into a product of genus one curves \cite{enolrev}. Special $K3$'s
 and complex multiplication also  appear in the attractor mechanism
\cite{moore,lps}.
 Whether the appearances of special $K3$'s in these two
 very different ways in string theory are related is an intriguing question.

\bigskip

{\bf Acknowledgements}:
We thank  Robert de Mello Koch, Jose Figueroa-O'Farrill, Pei Ming Ho,
Simon McNamara, Bill Spence, Radu Tatar, Steve Thomas, Nick Toumbas,
Gabriele Travaglini and  John Ward for discussions.
The work of SR is supported by 
a PPARC Advanced Fellowship. CP would like to acknowledge a
QMUL Research Studentship. This work was in
part supported by the EC Marie Curie Research Training Network
MRTN-CT-2004-512194.


\begin{appendix}

\section{Further aspects of space-time dependence of $S^2$ funnel}
\subsection{Lorentz Invariance of the BPS condition}
\par
We will study the supersymmetry of the 
space and time-dependent system considered in section 2,
in order to obtain a Lorentz invariant BPS condition.
 The vanishing of the
variation of the gaugino on the world-volume of the intersection will
require that  \cite{myers}
\begin{equation}\label{susy}
    \d\chi = \Gamma^{\m\n} F_{\m\n}\;\e=0\;,
\end{equation}
where $\m,\n$ spacetime indices, with
\begin{equation}
F_{ab}=0\qquad F_{ai}=D_{a}\Phi_{i}\qquad
F_{ij}=i[\Phi^{i},\Phi^{j}]\qquad\textrm{and}\qquad a=\s,\t\quad
,\quad i=1,2,3\;.
\end{equation}
Here the spinor $\e$ already satisfies the D-string projection $\Gamma^{\t\s}\e=\e$ and multiply indexed $\Gamma$'s denote their normalised, fully antisymmetric product, e.g. $\Gamma^{\m\n}=\frac{1}{2}[\Gamma^\m ,\Gamma^\n ]$. Then
\begin{equation}
     \left( i \G^{jk} [\Phi^{j},\Phi^{k}]+
2\G^{a i}D_{a}\Phi^{i}\right)\e=0\;.
\end{equation}
We will consider the conjugate spinor
\be
\overline{\d\chi}=\d\chi^{\dag}\;\Gamma^0
\ee
and construct the invariant quantity
\be\label{invariant}
\d\chi^\dag\;\Gamma^0\;\d\chi=0\;,
\ee
where our conventions for the $\Gamma$-matrices are
\be
(\Gamma^0)^2=-1\:,\quad (\Gamma^i)^2=1\quad\textrm{and}\quad (\Gamma^{\m})^\dag=\Gamma^0 \Gamma^\m \Gamma^0\;,\quad(\Gamma^{\m\n})^\dag =\Gamma^0 \Gamma^{\m\n}\Gamma^0\;.
\ee
Then (\ref{invariant}) becomes
\be\label{invariant2}
\e^\dag \left( -i \;(\Gamma^{\ell m})^\dag [\Phi^\ell ,\Phi^m ]+2 (\Gamma^{bi})^\dag D_b \Phi^i \right)\; \Gamma^0 \left(  i \;\Gamma^{jk} [\Phi^j ,\Phi^k ]+2 \Gamma^{ai} D_a \Phi^i \right)\e=0\;.
\ee
This will give three kinds of terms: Quadratic terms in derivatives, a quadratic term in commutators of $\Phi^i$ and cross-terms. The first of these will be
\be
4 (D_a \Phi^i)(D_b \Phi^j)\; \e^\dag (\Gamma^0 \Gamma^{bj}\Gamma^0) (\Gamma^0 \Gamma^{ai})\e\;.
\ee
By making use of the ansatz $\Phi^{i}=\hat{R}(\s,\t)\a^{i}$ and by taking a trace over the gauge indices this becomes
\be
4 (\pd_a \hat R)  (\pd_b \hat R)\;Tr\; (  \a^i \a^j )  \;\e^\dag \left(\Gamma^0 \frac{1}{2}\{\Gamma^b , \Gamma^a \}\Gamma^j \Gamma^i +\Gamma^0 \frac{1}{2}[\Gamma^b , \Gamma^a ]\Gamma^j \Gamma^i\right) \e
\ee
Since $\{\Gamma^\m ,\Gamma^\n \}=2 g^{\m\n}$, $Tr\; \a^i \a^j = \hat c\; \d^{ij}$ for a constant $ \hat c $ and the expression is  symmetric in $a,b$ we get
\be\label{der}
12\; \hat c\;  (\pd_a \hat R)(\pd^a \hat R)
 \left(\e^\dag \Gamma^0  \e \right)\;.
\ee
The product of commutators will be
\be
-[\Phi^\ell , \Phi^m] [\Phi^j , \Phi^k]\; \e^\dag \Gamma^0\Gamma^{\ell m} \Gamma^{jk}\e\;,
\ee
which with the trace and the use of the ansatz becomes
\be
4\hat c \hat R ^4 \; \varepsilon^{\ell m p}\varepsilon^{jkq}\d^{pq}\; \e^\dag \Gamma^0\Gamma^{\ell m} \Gamma^{jk}\e\;.
\ee
Since we are looking for a supersymmetry condition for the intersection, we want the spinor $\e$ to satisfy $\Gamma^{\t ijk}\e=\varepsilon^{ijk}\e$
 appropriate for  
 the $D3$-brane. Using this fact we can re-write the above as
\be
-4\hat c \hat R ^4 \; (\d^{\ell j}\d^{mk}-\d^{\ell k}\d^{m j})\;\varepsilon^{\ell m n}\varepsilon^{jkn}\; \e^\dag \Gamma^0 \e
\ee
and end up with
\be\label{comm}
-12\;\hat c (  4\hat R ^4 )  \;\left(\e^\dag \Gamma^0 \; \e\right)\;.
\ee
Finally, the cross-terms
\be
2i\left( \e^\dag \Gamma^0 \Gamma^{jk} \Gamma^{ai} [\Phi^j ,\Phi^k] (D_a \Phi^i)\e- \e^\dag \Gamma^0 \Gamma^{ai} \Gamma^{jk}  (D_a \Phi^i) [\Phi^j ,\Phi^k]\e\right)
\ee
will vanish  by following the same steps, since  $Tr\; \a^i \a^j \a^k\sim\varepsilon^{ijk}$ and the $\Gamma$-matrix commutator is $[\Gamma^i, \Gamma^{jk}]=(\d^{ij}\Gamma^k-\d^{ik} \Gamma^j)$. Collecting (\ref{der}) and (\ref{comm}) we have
\be
\e^\dag\;\Gamma^0 \e   \left( (\pd^a \hat R )(\pd_a \hat R) -4\; \hat R ^4 \right ) =0\;.
\nn
\ee
which implies 
\be 
 (\pd^a \hat R )(\pd_a \hat R) -4\; \hat R ^4  = 0 
\ee
By employing dimensionless variables, the supersymmetry-preserving condition is simply
\be\label{bpsboost}
(\pd_{\m}r)\;(\pd^{\m}r)=r^{4},\qquad \m=\s,\t \;.
\ee
Thus, every configuration that satisfies (\ref{bpsboost}) qualifies as a BPS state and
therefore will be semi-classically stable. The special case 
$ {\hat{ R}}^{\prime} = 2 ( \hat {R} )^2 $ was used in \cite{myers} and is 
related to Nahm's equations.
\par
We will briefly look at some solutions of this first-order equation. It is not hard to see
that the simplest ones  can be put into the form
\begin{equation}\label{boost}
    \hat{R}(\s,\t)=\pm \frac{1}{2}\frac{1}{\gamma (\s-\s_{\infty}+\b
    t)}\;,
\end{equation}
where $\s_{\infty}$ a constant, $\gamma=\frac{1}{\sqrt{1-\b^{2}}}$ and
$\b$ another constant, which can be thought of as the ratio of the
velocity of the system over the speed of light $c=1$. Then the
solution at hand is nothing but a boost of the static funnel
\be
\hat{R}= \pm\frac{1}{2(\s-\s_{\infty})}\;.
\ee
 Moreover (\ref{boost}) will
satisfy the full non-linear (\ref{nonlin}) and is therefore a
proper solution of our space and time-dependent system. In this case we can see explicitly how this solution  should also satisfy the YM equations of motion. By starting with the BPS-like relation (\ref{bpsboost}), one gets by differentiation and algebraic
manipulation
\begin{equation}
\pd_\m \pd^\n r\; (\pd_\n r) (\pd^\m r)=2 r^3 (\pd_\m r )(\pd^\m r) \;,
\end{equation}
which if substituted in (\ref{nonlinlor}) and again with (\ref{bpsboost}), will yield exactly $\partial_\m\partial^\m r=2r^{3}$, which are the YM equations.
Relations between the BPS condition, the DBI and 
 YM  equations were discussed in \cite{Brecher}.
 Therefore the class of space and time-dependent supersymmetric solutions to DBI \textit{simultaneously} satisfy the BPS and the
YM equations. Interestingly there exist solutions to one of YM or BPS which do not solve the DBI. For example, $r_{YM}=\frac{1}{\sqrt{2(\s^2-\t^2)}}$ solves the YM equations but not the BPS or DBI and  $r_{BPS}=\frac{1}{\sqrt{\s^2-\t^2}}$ is just a solution to BPS.
\par
The fact that we have recovered, in the context of space
\textit{and} time dependent transverse scalars, a boosted static
solution is something that we should have expected: The expression
for the Born-Infeld action is, of course, Lorentz invariant. This
guarantees that its extrema, and solutions to the equations of
motion, although transforming accordingly under boosts, should
still be valid solutions at the new points $x'=\Lambda^{-1} x$. The
lowest energy configuration of the system is naturally the BPS one
and a boost provides a generalisation that is still stable and
time-dependent. A natural consequence of this is the boosted brane array, which is indeed a solution of the space and time dependent DBI equations of motion
\begin{equation}
r(\s,\t)=\pm r_{0}\frac{1}{Cn\left(\frac{\sqrt{2} r_{0} \gamma(\s-\b\t)}{\sqrt{r_{0}^{4}+1}},\frac{1}{\sqrt{2}}\right)}
\end{equation}
and the boosted collapsing brane
\be
r(\s,\t)=\pm r_{0}Cn\left(\frac{\sqrt{2} r_{0} \gamma(\t-\b\s)}{\sqrt{r_{0}^{4}+1}},\frac{1}{\sqrt{2}}\right)\;.
\ee

\subsection{ Chern-Simons  terms and D-brane  charges for D1-D3 system }
\par
We found in section 2.1 that the equations of motion for a space and time-dependent `funnel' configuration have a solution both in the D1 and  the D3 picture. The Chern-Simons part of the non-abelian DBI in a flat background, is not relevant in the search for the equations of motion. However, it would be a nice check to see whether the charge calculation also agrees when time dependence of the scalars is introduced, as happens for the static case. We
will be assuming spherical symmetry of the solutions in what follows.
\par
We expect to recover a coupling to a higher dimensional brane through the non-commutative, transverse scalars and the dielectric effect. This is indeed the case and one gets for the $D3$-charge of the $D1$ configuration for \textit{any} time-dependent solution, at large-$N$
\begin{equation}\label{d1d3}
S_{cs}^{D1}=\frac{2\m_{1}}{\lambda}\int d\t\wedge d\s \;R_{ph}^2 \left( C^{(4)}_{\s123}\dot{R}_{ph}-C^{(4)}_{\t123}R'_{ph} \right) \;.
\end{equation}

The Chern-Simons part of the low energy effective D3-brane action reads
\begin{equation}
S_{CS}=\m_{3}\int \;Str\;P [C^{(4)}]+\m_{3}\int\;Str\;
P[C^{(2)}]\wedge F+\m_{3}\int\;Str\; P[C^{(0)}]\wedge F\wedge F\;.
\end{equation}
By considering a spherical co-ordinate embedding in the static gauge, the calculation of the D3 charge will give
\begin{equation}\label{d3d3}
S_{CS}^{D3}=4\pi\m_{3}\int dt\wedge dR_{D3}\;R_{D3}^2 \left[C^{(4)}_{t123}+C^{(4)}_{9123}\lambda\dot{\Phi}\right]\;,
\end{equation}
where $R_{D3},t$ are world-volume indices and $1,2,3$ are space-time indices.
\par
Following \cite{nonabelian} we will make the following identifications
\begin{itemize}
\item The physical radius of the fuzzy $S^{2}$ on the $D1$ side should correspond to the co-ordinate $R_{D3}$ on the world-volume
of the D3
\begin{equation}
R_{ph}\longleftrightarrow R_{D3}\;.
\end{equation}
\item The $\s$ co-ordinate on the world-volume of the $D1$ should be analogous to the transverse scalar $\Phi$ on the $D3$
\begin{equation}
\s\longleftrightarrow  \lambda\Phi\;.
\end{equation}
\item We finally assume
\begin{equation}\label{time}
\t\longleftrightarrow  t\;.
\end{equation}
\end{itemize}
\par
By thinking about the physical radius of the fuzzy sphere as a function of $\s$ and $t$, using the relationships (\ref{rphi}) and also the fact $(2\p\sqrt{\a'})^{p'-p}\m^{p'}=\m^p$, we are able to write (\ref{d3d3}) in terms of $D1$
world-volume quantities. Everything works out nicely and one gets
\begin{equation}
S_{CS}^{D1}=S_{CS}^{D3}\;.
\end{equation}\label{ansatz}
Furthermore we can make use of  (\ref{boost}) to get:
\begin{equation}\label{ansatzboost}
    \Phi(R_{D3},t)=\pm\left(\frac{N}{2\gamma R_{D3}}-\frac{\b t}{\lambda}\right)
\end{equation}
as the boosted solution on the $D3$. This can be re-written as
\be
\lambda\Phi(R_{D3},t)=\gamma(\lambda\Phi_{0}(R_{D3})+\b t')\;,
\ee
where $t'$ is related to $t$ by a Lorentz transformation $t'=\gamma(t-\b\lambda\Phi)$ and can be viewed as the time co-ordinate for an observer on the worldvolume of the boosted $D3$.

\section{ Lagrangians for holomorphic differentials }

We have seen that the space or time-dependence
of the radius of the fuzzy $S^4$ described by the Lagrangian
(\ref{d15}) is given  by
integrating the holomorphic differential $ { dr \over  s }  $
on the curve $(r,s)$ given by (\ref{eqwrts}).  
There are  more general holomorphic differentials
on the same Riemann surface, which enter
on equal footing in the geometry of the Riemann surface.
For the hyperelliptic curves of the type we considered,
they are of the form $ { r^{\alpha } dr \over s } \equiv dt_{\alpha}  $,
where $\alpha $ can take values from $1$ to the genus of the curve.  
It is natural to ask if there are Lagrangians such that the 
time elapsed or distance are given by the more general 
holomorphic differentials. This is indeed possible and the Lagrangian 
densities are 
\be
L_{\alpha} 
 = - \sqrt{ 1 - r^{2 \alpha}  ( {\dot r}^2 + {\rp}^2 )   } ( 1 + r^4 )\;.
\label{lagsfrhol}
\ee

The automorphisms we discussed in section 4, 
can be used to relate the time evolution of interest 
which appears as the imaginary part of $u_1$, 
to the space evolution for $ \alpha = 2 $. The variable 
$u_2 $ discussed in section 4.2 and 4.3 are related to 
the $ \alpha =2 $ holomorphic differential of the genus three 
curve and hence to the $ \alpha = 2 $ Lagrangian above.

It is also  interesting to explore how the second order 
equations following from $L_{\alpha}$ transform under inversion of $r$. 
 Dropping the space dependence, the equation of motion 
following from (\ref{lagsfrhol}) is
\be 
\ddot r = - { 4 r^{3 - 2 \alpha } \over { 1 + r^4 } } +
{ { \dot r}^2 \over r ( 1 + r^4 ) }
\bigl (~ ( 4 - \alpha )r^4 - \alpha ~ \bigr )    \;.
\label{lagholeom}
\ee 
Under a transformation $ R = { 1 \over r } , \tau = t $,
we get 
\be 
\ddot R =  { 4 R^{ 3 + 2 \alpha } \over 1 + R^4 }  +
{ {\dot R}^2 \over R ( 1 + R^4 ) }
\bigl( ~  ( \alpha + 2  ) R^4  + ( \alpha - 2 ) ~ \bigr ) \;.
\label{tlagholeom}
\ee 
If we ignore, for the moment,  the first term on the RHS, 
we see that the  $ \alpha = 0 $ equation   in (\ref{lagholeom}) 
maps to the $ \alpha = 2 $ term. 
If we neglect the  velocity dependent term on the RHS,
the $ \alpha = 0 $ maps back to $ \alpha = 0 $. 

Further consider the transformation $ R  = { 1 \over r } ,
{ d \tau \over d t }  = -{  1 \over R^2 }  $, to get
\be 
\ddot R = { 4 R^{ 7 + 2 \alpha } \over 1 + R^4 }
 + { {\dot R}^2 \over R ( 1 + R^4 ) }
\bigl( ~  ( \alpha + 4  ) R^4  +  \alpha  ~ \bigr ) \;.
\label{ttlagholeom}
\ee 
If we keep only the velocity dependent term on 
the RHS, we find that the case $ \alpha = 0 $ from (\ref{lagholeom})
maps to the case $ \alpha =0 $ of  (\ref{ttlagholeom}). 

While these transformations include  relations between 
the $ \alpha = 0 $ and $ \alpha =2 $ Lagrangians, they require 
neglecting some terms in the equation of motion so they would hold 
in special regimes where these are valid.
 The transformations involving $r_0$, which we described in section 4,
   give a more direct  relation between the functions
solving equations from $L_0 $ and $L_2$.  It will be interesting
to find physical brane systems which directly lead to $L_{\alpha } $ for
$ \alpha \ne 0$.

\section{Evaluation of the solution of the collapsing $D2$ configuration using Weierstrass $\wp$-functions }
\par
Here is a warm-up which makes use of the technology employed for the solution of the Jacobi inversion problem, in terms of Weierstrass $\wp$-functions. We use this to recover the familiar result from the 2-sphere collapse.
\par
The integral related
to the time of  collapse from an initial radius $r_{0}$ and in
dimensionless variables is given by the expression
 \be
 \int_{r_{0}}^{\rp}\frac{\sqrt{1+r_{0}^{4}}}{\sqrt{r_{0}^{4}-r^{4}}}dr=-t\;.
\ee
After a re-scaling $\tilde{r}=\frac{r}{r_{0}}$ we have
\be
  \int_{1}^{\tilde{r}^{\prime}}\frac{\sqrt{1+r_{0}^{4}}}{r_{0}\sqrt{1-\tilde{r}^{4}}}d\tilde{r}=-t
\ee
and
\be
   \int_{1}^{\tilde{r}^{\prime}}\frac{d\tilde{r}}{\sqrt{\tilde{r}^{4}-1}}=-i\frac{r_{0}t}{\sqrt{1+r_{0}^{4}}}=u_1\;.
\ee
Next we will perform the substitution $\tilde{r}^{2}=x$ ending up with
\be
    u_1=\int_{1}^{x^{\prime}}\frac{dx}{\sqrt{4x(x^{2}-1)}}.
\ee
This is a definite integral over the holomorphic differential of the elliptic curve
\be
y^2=4x^3-4x\;,
\ee
with branch points at $x={0,\pm1}$ and infinity.
\par
It is easy to see that the former integral can be decomposed into the following
two
\be\label{twints}
     u_1=\int_{1}^{\infty}\frac{dx}{\sqrt{4x(x^{2}-1)}}-
\int_{x^{\prime}}^{\infty}\frac{dx}{\sqrt{4x(x^{2}-1)}}\;,
\ee
which are exactly the real half-period $\Omega$ of the surface
and the inverse of the Weierstrass elliptic function
$\wp^{-1}(x^{\prime},g_{2},g_{3})$ with invariants
$g_{2}=4,g_{3}=0$. Both the real and
the imaginary half-period $(\Omega,\Omega')$ can be calculated by contour
integration after we have defined a homology basis on the surface.
In this case take the $a$-cycle to be the loop surrounding the
points $(1,\infty)$ (or equivalently, by deformation, the points
$(-1,0)$) and the $b$-cycle the loop around $(0,1)$ and across the
two sheets. The results are
 \be
   \Omega=\int_{0}^{1}\frac{ds}{\sqrt{4s(1-s^{2})}}\qquad\textrm{and}
\qquad\Omega'=i\int_{0}^{1}\frac{ds}{\sqrt{4s(1-s^{2})}}\;,
\ee
with $s$ a real, positive integration parameter. The modulus of the torus is then simply given by $\t=\frac{\Omega'}{\Omega}=i$.
 There exists an alternative definition for this,
namely $\t=i\frac{K(k)}{K(k')}$, where $K(k)$ is the complete elliptic
 integral of the first kind and $k,k'$ the elliptic
modulus and the complementary modulus respectively, with
$k^{2}+k'^{2}=1$. This implies $K(k)=K(k')$ and $k=k'=1/\sqrt{2}$
for the functions and integrals defined on this surface with this
 particular choice of cycles and cuts.
\par
 From the definition of $\wp(u)$ one can derive several relations
 between the  latter and the Jacobi elliptic
functions \cite{byrd}. Take a general curve
\be
  \wp^{-1}(w)\equiv v
  =\int_{w}^{\infty}\frac{dz}{\sqrt{4z^{3}-g_{2}z-g_{3}}}=\int_{w}^{\infty}
\frac{dz}{\sqrt{4(z-e_{1})(z-e_{2})(z-e_{3})}}
\ee
and by setting
\be\label{es}
\g^{2}=e_{1}-e_{3}\;,\quad k^{2}=\frac{e_{2}-e_{3}}{e_{1}-e_{3}}\quad
\textrm{and}\quad z=e_{3}+\frac{\g^{2}}{s^{2}}\;,\quad
\textrm{for}\quad e_{1}>e_{2}>e_{3}\;,
\ee
 the integral becomes
  \be
   v=\frac{1}{\g}\int_{0}^{W}\frac{ds}{\sqrt{(1-s^{2})(1-k^{2}s^{2})}}\;.
\ee
This is simply the definition for the Jacobi elliptic integral of the
first kind and is invertible with $Sn^{-1}(W)=\g v $. If we use
this fact and (\ref{es}) we get
\be
\wp(v)=e_{3} + \frac{\g^{2}}{Sn^{2}(\g v,k)}
\ee
and for the example at hand, $e_{1}=1, e_{2}=0, e_{3}=-1$ and
the relationship becomes
\be\label{psn}
\wp(v)=-1+\frac{2}{Sn^{2}(\sqrt{2}v,\frac{1}{\sqrt{2}})}\;.
\ee
Returning to (\ref{twints}) we have
\be
u_1 = \Omega-\wp^{-1}(x';4,0)
\ee
and
\bea \label{subs}
\nn x'&=& \wp( u_1 -\Omega;4,0)\\
  &=& 1+\frac{2}{\wp( u_1 ;4,0)-1}\;,
\eea
where we have made use of the identity
\be
\wp(v\pm\Omega)=e_{1}+\frac{(e_{1}-e_{2})(e_{1}-e_{3})}{\wp(v)-e_{1}}\;,
\ee
Now from (\ref{psn}) and by using the following properties of elliptic functions
\be
 Cn( v , k ) = { 1 \over Cn (iv , k' ) } \qquad \textrm{and}
 \qquad   Sn^{2}(v,k)+Cn^{2}(v,k)=1\;.
\ee
 we obtain
\bea
\nn x'&=&  Cn^{2}\left(\sqrt{2}i u_1,\frac{1}{\sqrt{2}}\right)\;.
\eea
 Converting back to the original quantities $x'=\tilde{r}'^{2}=r'^{2}/r_{0}^{2}$  and by substituting
 $u_1=-ir_{0}t/\sqrt{1+r_{0}^{4}}$, we
recover the desired result
\be
r'=r_{0}\;Cn\left(\frac{r_{0}\sqrt{2}t}{\sqrt{1+r_{0}^{4}}},
\frac{1}{\sqrt{2}}\right)\;.
\ee

\end{appendix}



\end{document}